\documentclass[hidelinks,onefignum,onetabnum]{siamart250211}

\usepackage{lipsum}
\usepackage{amsfonts}
\usepackage{graphicx}
\usepackage{epstopdf}
\usepackage{algorithmic}
\ifpdf 
  \DeclareGraphicsExtensions{.eps,.pdf,.png,.jpg}
\else 
  \DeclareGraphicsExtensions{.eps}
\fi

\newsiamremark{remark}{Remark}
\newsiamremark{hypothesis}{Hypothesis}
\crefname{hypothesis}{Hypothesis}{Hypotheses}
\newsiamthm{claim}{Claim}
\newsiamremark{fact}{Fact}
\crefname{fact}{Fact}{Facts}

\headers{A Cycle Walk for Sampling Spanning Forest Measures}{D. DeFord, G. Herschlag and  J. Mattingly}

\title{A Cycle Walk for Sampling Measures \\on Spanning Forests for Redistricting}

\author{Daryl R. DeFord\thanks{
    Mathematics and Statistics Department, Vassar College, Poughkeepsie,  NY 12604, USA (\email{ddeford@vassar.edu})}
  \and Gregory Herschlag\thanks{Mathematics Department, Duke University, Durham, NC 27710, USA (\email{gregory.herschlag@duke.edu})}
  \and Jonathan C. Mattingly 
\thanks{Mathematics Department and Department of Statistical Science, Duke University, Durham, NC 27710, USA 
  (\email{jonathan.mattingly@duke.edu})}
}

\usepackage{amsopn}

\usepackage{graphicx} 
\usepackage{float}
\usepackage{subcaption}
\usepackage{caption}
\usepackage{url}
\usepackage{hyperref}
\usepackage{cleveref}
\usepackage{lmodern}

\usepackage{subcaption}

\newcommand{\downUp}{\textrm{upDown}}
\newcommand{\treeOne}{\textrm{1-tree }}
\newcommand{\treeTwo}{\textrm{2-tree }}
\newcommand{\eqdef}{\overset{{\rm def}}{=}}

\newcommand{\R}{\mathbf{R}}

\newcommand{\X}{\mathbf{X}}

\newcommand{\Tree}{\mathrm{tree}}
\newcommand{\Compact}{\mathrm{compact}}
\newcommand{\ccdot}{\;\cdot\;}
\usepackage{xcolor}  

\definecolor{customgray}{RGB}{200,200,200}

\usepackage{todonotes}
\usetikzlibrary{decorations.pathmorphing}

\ifpdf
\hypersetup{
  pdftitle={A Cycle Walk for Sampling Measures on Spanning Forests for Redistricting},
  pdfauthor={D. DeFord, G. Herschlag and  J. Mattingly}
}
\fi

\begin{document}
\maketitle

\begin{abstract}
We introduce the \emph{Cycle Walk}, a new Markov chain Monte Carlo method for sampling distributions on balanced graph partitions, motivated by applications in political redistricting. The method operates on spanning forests and combines two types of updates: local “cycle” moves within districts and global moves that exchange population between adjacent districts while preserving balance constraints. This construction enables efficient Metropolis--Hastings correction while allowing proposals at multiple spatial scales.

We show that the Cycle Walk naturally interpolates between existing approaches based on local updates and a class of global update methods derived from recombination (RECOM). Through a range of numerical experiments on synthetic graphs and real-world precinct data, we demonstrate that the Cycle Walk exhibits improved empirical convergence diagnostics for distributions that place weaker weight on spanning-tree counts, a regime that is challenging for existing methods. In particular, the algorithm remains effective when incorporating alternative compactness measures that more closely reflect policy-relevant criteria.

These results suggest that the Cycle Walk provides a flexible and computationally efficient framework for sampling from a broader class of redistricting distributions than previously accessible with MCMC techniques.
  \end{abstract}



\section{Introduction}

We are motivated by the problem of sampling redistricting plans respecting specified policy or legal constraints and preferences. The use of a large ensemble of redistricting plans to evaluate a particular plan has become standard in recent legal and policy discussions. As such, it has become critical to sample redistricting plans that incorporate increasingly nuanced policy and legal considerations.  A number of methods have been proposed to perform such sampling. While there have been successes in sampling \cite{MattinglyVaughn2014,Bangia17,jcmReport,deford2019recombination,deford2019redistricting,herschlag2020quantifying,jcmReportHarperVHallMoore,duchinPAreport,autry2021metropolized,autry2023metropolized,autry2020multiscale,zhao2022mathematically},  the ability to sample ensembles satisfying many natural policy-motivated specifications remains largely out of reach
\cite{MattinglyVaughn2014, herschlag2020quantifying, duchinPAreport, zhao2022mathematically,jonasBlogPost}. The post in \cite{jonasBlogPost} highlights biases that may arise when using less transparent measures.

We propose a new class of Markov chains on the space of redistricting plans based on creating and splitting cycles. It is related to the foundational theoretical work on the Up-Down Walk on spanning trees and forests analyzed in \cite{AnariVinzantVuong2021, russo2018linking}, but is designed to produce partitions with balanced populations, a requirement in redistricting. Initial tests indicate that it can successfully sample from important examples that have been challenging for existing methods.  In particular, we expect this new class of Cycle Walk samplers to be more successful in sampling from the increasingly nuanced sampling specification that future redistricting investigations will require.  Lastly, we comment that the Cycle Walk sampler has been very fast in our preliminary experiments.

\section{Setup}

We will encode the region to be redistricted by a planar graph $G=(V,E)$ whose vertices $V$ represent the geographic units out of which districts are built and the edges $E$ encode which geographic units are adjacent. Often, the region to be redistricted will be a state and geographic units will be either precincts or census blocks.

We will cast the problem of finding a redistricting plan with $d$ districts as that of finding a partition of the vertex set $V$ into $d$ sets. One convenient representation of a redistricting plan with $d$ districts is a coloring of the graph with $d$ colors, which are represented by the numbers $1,\dots,d$. The coloring is a function $\xi:V \rightarrow \{1,\dots,d\}$. With this notation, the $d$ districts $D_1, \dots, D_d$ are defined by
\begin{align}\label{eq:district_def}
  D_i(\xi)=\big\{ v \in V : \xi(v)=i\big\}
\end{align}
If $\xi$ is clear in the context, we will at times write $D_i$. If there are two plans, $\xi$ and $\xi'$, we will at times abbreviate the $i$th district in each plan as $D_i$ and $D_i'$, respectively.

A convenient way to specify the collection of maps of interest is to place a probability measure, $\pi$, on the space of redistricting plans, which we will denote by $\mathcal{D}$. In the redistricting context, this dates back at least to \cite{MattinglyVaughn2014,fifield2015}.  We will specify the measure by placing  a score function $J\colon \mathcal{D} \rightarrow (-\infty,\infty]$ and then defining the target distribution $\pi$ by 
\begin{align}\label{eq:pi_def}
   \pi(\xi) \propto e^{-J(\xi)}
\end{align}
Since for a finite graph $G$ the space $\mathcal{D}$ contains a finite number of elements, the expression in \eqref{eq:pi_def} can be normalized to define a probability measure. The assignment $\xi$ defines a partition, with $d$ elements, of the vertex set $V$.  We will denote the space of all such partitions by $\mathcal{P}_d(V)$. We assume that $J(\xi)$ is invariant under permutations of the labels. Namely if $\sigma$ is a permutation of the numbers $\{1,\cdots,d\}$ then $J(\sigma(\xi))=J(\xi)$. Thus $\pi$ is a measure on $\mathcal{P}_d(V)$.

For mathematical and design reasons, one often takes the score function $J$ to be of the form
\begin{align*}
  J(\xi)=J_1(\xi)+\cdots+J_k(\xi)\;.
\end{align*}
Each of the $J_i$ might encode our preferences around a particular property. Preferences may take the form of a hard constraint -- a condition that must be satisfied by every member of our ensemble -- or as soft constraints -- properties that we place higher probability on but do not strictly enforce. An example of a hard constraint is the requirement that every district  $D_i(\xi)$ from \eqref{eq:district_def} of a given plan $\xi$ is pathwise connected. In this case, one could encode this constraint by defining
\begin{align*}
  J_{\text{connected}}(\xi)=
  \begin{cases}
    0  & \text{ if all $D_i(\xi)$ are  pathwise connected}\\
    +\infty & \text{ Otherwise}
  \end{cases}\;.
\end{align*}

In some contexts, there is also a legal requirement that all districts contain a population that deviates from the ideal population by no more than a specified fraction $\delta_{\text{population}}$. There are many approaches to handle this balance condition. One could again use a hard constraint by requiring that  $\sup_i |\textrm{PopDev}(D_i(\xi))| \leq \delta_{population}$ where $\textrm{PopDev}(D_i(\xi))$ is the population deviation of the district $D_i(\xi)$ and $\delta_{population}>0$ is desired tolerance for population deviation. One would then set $J_{\textrm{population}}(\xi)=0$ when this bound is satisfied and $+\infty$ when it is not. Alternatively, one might use  a soft constraint via a penalization score function
\begin{align*}
 J_{\textrm{population}}(\xi) = \phi\Big( \sum_i \big|\textrm{PopDev}(D_i\big(\xi)\big)\big|\Big) 
\end{align*}
where $\phi$ is a non-decreasing function. One may also combine soft penalization with a hard threshold and use a score function of the form
\begin{align*}
  J_{\textrm{population}}(\xi) =
  \begin{cases}
  \displaystyle  \phi\Big( \sum_i \big|\textrm{PopDev}(D_i\big(\xi)\big)\big|\Big) 
& \text{ if $ \displaystyle \sup_i   |\textrm{PopDev}(D_i(\xi))| \leq \delta_{\text{population}} $}\\
   \displaystyle   +\infty & \text{ if $\displaystyle \sup_i   |\textrm{PopDev}(D_i(\xi))| > \delta_{\text{population}} $}
  \end{cases}
\end{align*}
Another common requirement is that districts be compact. This can be achieved again by a soft penalization of some compactness metric (e.g., average isoperimetric score of the districts) or a mixture of soft and hard thresholding. As discussed below, it is sometimes algorithmically convenient to penalize on the sum of the log of the number of spanning trees on each district \cite{autry2023metropolized,autry2021metropolized,cannon2022spanning} as this is correlated with certain definitions of compactness \cite{deford2019recombination}; however, this has some complications (e.g., \cite{jonasBlogPost}) and may not align with policymakers' preferences.

\section{Motivating Algorithms} 

We briefly recall several methods that are useful for understanding the development of the Cycle Walk. We begin with the single-node-flip walk.

\subsubsection*{The Single-Node-Flip Walk}

The single-node-flip Markov chain on partitions evolves by randomly selecting two adjacent districts and a boundary vertex between them according to a specified probability distribution. The selected vertex is then reassigned from its current district to the other chosen district.
There are many ways to select the pair of adjacent districts and the vertex to be reassigned. For example, one may first choose a pair of adjacent districts according to some rule (e.g., uniformly at random over all adjacent pairs), then select an edge along their shared boundary, and finally choose one of the endpoints of that edge to reassign.
If the current plan is $\xi$, this procedure defines a Markov chain with transition probabilities $Q(\xi,\xi')$, representing the probability of moving from $\xi$ to $\xi'$ in one step. Under mild assumptions on the graph $G$, the single-node-flip walk has a unique stationary distribution $\pi_{\mathrm{SNF}}$, and if $\xi_{n+1}$ is drawn from $Q(\xi_n,\cdot)$, then the distribution of $\xi_n$ converges to $\pi_{\mathrm{SNF}}$ as $n \to \infty$.
However, for a general planar graph $G$ and score function $J$, this stationary distribution $\pi_{\mathrm{SNF}}$ does not coincide with the desired target distribution $\pi$. This discrepancy can be addressed using the Metropolis--Hastings procedure.

\subsubsection*{Metropolis-Hastings}
The Metropolis--Hastings algorithm modifies a Markov transition kernel by introducing rejections in order to target a prescribed stationary distribution. 
Given a Markov transition kernel $Q$, the following procedure defines a new transition kernel $P$ with invariant distribution $\pi$.
If the current state is $\xi_n$, we \textit{propose} a new state $\xi'$ according to $Q(\xi_n,\cdot)$. We then \textit{accept} this proposal and set $\xi_{n+1} = \xi'$ with probability
\begin{align*}
a(\xi,\xi') = 1 \wedge \frac{\pi(\xi') Q(\xi',\xi)}{\pi(\xi) Q(\xi,\xi')}.
\end{align*}
Otherwise, the proposal is rejected and the chain remains at $\xi_n$.
We refer to $Q$ as the proposal kernel, $a$ as the acceptance probability, and $P$ as the Metropolization of $Q$. Computing the acceptance probability requires evaluating both the forward transition probability $Q(\xi,\xi')$ and the reverse probability $Q(\xi',\xi)$.
Under mild assumptions on $Q$ and $\pi$, the Markov chain defined by $P$ converges to a unique stationary distribution $\pi$. Using the single-node-flip walk as the proposal kernel has the advantage that it is computationally efficient and easy to implement. However, in practice, this approach has several drawbacks.

\subsubsection*{Recombination}

In theory, the Metropolized single-node-flip Markov chain described above can sample from any reasonable measure $\pi$ that satisfies appropriate irreducibility requirements. However, convergence may be prohibitively slow.  Part of the difficulty likely stems from the fact that single-node-flip only proposes local, small-scale changes to the district boundaries. Another problem is that the proposal tends to both unbalance the population and generate less compact districts, which can lead to a high rejection rate of the proposed moves. Even ignoring these issues, energetic barriers may remain \cite{herschlag2020nonreversiblemarkovchainmonte,chuang2024multiscaleparalleltemperingfast}.

The Recombination Markov chain (RECOM) introduced in \cite{deford2019recombination} addresses these issues. It can consistently keep the population balanced while performing global, large-scale moves close to or equal to the scale of an entire district. The algorithm selects two adjacent districts according to some specified distribution, merges the districts into a single region, and then splits this region into two new districts. The splitting process naturally produces population-balanced districts. 

To divide the merged region, one draws a random spanning tree on the induced graph of the merged region. Removing an edge from a spanning tree always produces two trees whose vertices partition the sub-region into two. By traversing the spanning tree, one can then efficiently identify the edges that, if removed, would split the merged region into two new districts with balanced populations within the specified population tolerance. Then, a random edge is chosen to remove from among those identified.

For concreteness, one might choose the adjacent districts uniformly among all adjacent pairs. Similarly, one may choose uniformly from the edges on the merged spanning tree that produces two balanced regions within the tolerance. In the original formulation, the RECOM algorithm chose the random spanning tree uniformly among all spanning trees on the induced graph of the merged region. This can be done reasonably efficiently using Wilson's algorithm. In some recent implementations of RECOM, the spanning tree is drawn from the random minimal spanning tree distribution, which can be very efficiently generated by using Kruskal's algorithm with random edge weights.

A key strength of RECOM is that it performs large, global moves, leading to reduced dependence between successive states. However, the stationary distribution of RECOM is not known in general, and different implementations yield different distributions. In particular, RECOM does not sample from the target distribution $\pi$ defined in \eqref{eq:pi_def}.

In principle, one could Metropolize RECOM, however, calculating the reverse (or backward) probabilities turns out to be computationally infeasible. In \cite{autry2023metropolized,autry2021metropolized}, the authors modified RECOM by lifting it to the space of a spanning forest and showing that this makes the computation of the reverse probabilities computationally feasible.  In \cite{cannon2022spanning}, the authors present a different solution: They modify RECOM to satisfy detailed balance with respect to the uniform measure on spanning trees. This also makes Metropolization practical. We will describe the first approach as it will inform the ideas of the Cycle Walk.

\subsubsection*{Metropolized Forest Recombination} Metropolized Forest Recombination \cite{autry2023metropolized} can be understood as a lifting of RECOM to the space of spanning forests; on forests, the proposal chain can be efficiently Metropolized. Alternatively, it can be viewed as a form of MCMC with data augmentation to make Metropolization computationally feasible. While RECOM naturally evolved on the space of partitions of the vertex set $V$ into $d$ districts, Metropolized Forest Recombination is most naturally described on the space of spanning forests with $d$ spanning trees. Recall that a spanning forest is a collection of disjoint trees on $G$ so that the union of the vertices in the trees is $V$. We will write $\mathcal{F}_d$ for the space of spanning forests with exactly $d$ trees. 

We proceed in a way that closely mirrors the RECOM algorithm. Given a $\tau \in \mathcal{F}_d$ we obtain a new random $\tau' \in \mathcal{F}_d$ by randomly choosing two trees in $\tau$ that are adjacent according to some rule. We then draw a new spanning tree randomly on the graph induced by merging the vertices of the two chosen trees.  Next, we split this randomly chosen spanning tree into two trees by removing a randomly chosen edge from those whose removal yields two trees satisfying the population bounds.  The spanning forest $\tau'$ is then defined as $\tau$ with the two initially chosen trees replaced with the two new trees obtained by splitting the random spanning tree in two by removing an edge. 

As with RECOM, there are many choices for how to choose the new random spanning tree and how to choose which edge to remove to obtain two relatively balanced trees. It is shown in \cite{autry2023metropolized} that the probability of transition from $\tau \rightarrow \tau'$ and from $\tau' \rightarrow \tau$ can efficiently be calculated. This allows the Forest Recombination walk to be efficiently Metropolized.  Since the initial step in this algorithm is to draw a new random spanning tree, we can equally view Metropolized Forest Recombination as a Markov chain on the space of partitions by simply projecting $\tau'$ down to the partition it represents. However, it is important to Metropolize the walk before this projection if that is desired. An alternative method extends the space of spanning forests to a linked-spanning forest in which each pair of adjacent trees has a unique spanning edge \cite{autry2021metropolized}; we will refer to this method as linked-Forest Recombination. In practice, the two methods are similar but have a slight difference in their implementation.

\section{The Cycle Walk}
\label{sec:cycleWalk}
We now present the Cycle Walk on balanced partitions, which is the main topic of this note. We will fix some algorithmic choices for clarity; at the end, we will comment on possible variations and extensions.

We will begin by formally introducing our setting and notation, and fix a finite undirected graph $G=(V,E)$.
Given any subgraph $G' \subset G$, we will let $V(G')$ and $E(G')$ be respectively the vertex and edge set of $G'$.  If $G',G'' \subset G$, then $G'\cup G''$ is the graph induced in $G$ by the vertices in $V(G')\cup V(G'')$. Given a subgraph $G'=(V',E')$, we define $\partial G'$ by $\partial G'=\{ e=(u,v) \in E:  e \not \in E', u \in V'\}$.  If $V' \subset V$, then the graph $G'=(V',E')$  induced by $V'$ is defined by $E'=\{ e=(u,v) \in E : u,v \in V'  \}$.

A spanning forest $\tau=\{t_1,\dots,t_d\}$ on $G$ is a collection of  trees $t_1,\dots,t_d$ in $G$ with $V(t_i)\cap V(t_j)=\emptyset$ for $i \neq j$ and $\cup_i V(t_i)=V$. We consider $\tau$ to be an unordered set of trees.  We will write $|\tau|$ for the number of trees in the spanning forest $\tau$.  We will let $\mathcal{F}(G)$ be the space of all spanning forests on $G$ and 
\begin{align}\label{eq:Fd}
  \mathcal{F}_d(G) = \Big\{  \tau  \in \mathcal{F}(G) : |\tau|=d \Big\}
\end{align}
$\mathcal{F}_1(G)$ is just the space of spanning trees on the graph $G$. Observe $\tau=\{t_1,\dots,t_d\} \in  \mathcal{F}_d(G)$ defines a unique partition by the assignment $\xi_\tau(v)=i$ if $v \in t_i$. Similarly for a partition assignment $\xi \in \mathcal{P}_d(G)$, we define the set of all spanning forests $\tau_\xi \subset \mathcal{F}_d(G)$ compatible with the partition $\xi$ by
\begin{align*}
  \tau_\xi=\{ \tau' \in \mathcal{F}_d(G)  : \xi_{\tau'} = \xi \}
\end{align*}
and the number of forests corresponding to the assignments $\xi$
\begin{align}
\Tree(\xi)=| \tau_\xi|\,.
\end{align}
The number of forests $\Tree(\xi)$ compatible with  a partition $\xi=\{D_1,\dots,D_d\}$ can be calculated by multiplying the number of spanning trees on each element of the partition. Namely,
\begin{align}
\Tree(\xi)=\prod_{i=1}^d \Tree(D_i(\xi)) \qquad \text{where}\qquad  \Tree(D_i) =| \tau_{D_i}|\,,
\end{align}
where
\begin{align*}
  \tau_{D_i}=\{ t \in \mathcal{F}_1(D_i, E(D_i))\} ,
\end{align*}
and where $E(D_i(\xi))$ are the edges of the graph connecting two vertices in $D_i(\xi)$. Hence $(D_i, E(D_i))$ is the subgraph induced by the vertex set $D_i$.

There are many natural ways to lift  $\pi(\xi)$, a measure on partitions (see \eqref{eq:pi_def}), to a measure on forests. 
Two natural choices are
\begin{align}
  \nu_0(\tau) \propto \pi(\xi_\tau) \qquad\text{and}\qquad  \nu_1(\tau) \propto \frac{\pi(\xi_\tau)}{\Tree(\xi_\tau)}\label{eq:nu_0_1}
\end{align}
Notice that the projection of $\nu_1$ to a measure on $\mathcal{P}_d(G)$ agrees with $\pi$ since $\nu_1(\tau_\xi)= \pi(\xi)$. In other words,  $\nu_1(\tau_\xi)$ is the probability  of drawing a forest representing the partition $\xi$. More concretely,
\begin{align*}
    \nu_1(\tau_\xi) = \sum_{\tau \in \tau_\xi} \nu_1(\tau) = \frac{\pi(\xi_\tau)}{\Tree(\xi_\tau)} \sum_{\tau\in \tau_\xi} 1 =  \pi(\xi).
\end{align*}
Furthermore, if $J \equiv 0$ then $\nu_0(\tau)$ is uniform on the space of  spanning forests $\mathcal{F}_d(G)$ while $\nu_1(\tau)$ is the lift to $\mathcal{F}_d(G)$ of  the uniform measure on the space of partitions  $\mathcal{P}_d(G)$. We may interpolate between $\nu_0$ and $\nu_1$ via
\begin{align}
\nu_{\gamma}(\tau) \propto \frac{\pi(\xi_\tau)}{\Tree(\xi_\tau)^{\gamma}},
\label{eq:nudefinition}
\end{align} 
where we consider $\gamma\in[0,1]$ as an interpolating parameter between the two natural choices. One could also consider more generic choices of $\gamma \in \mathbf{R}$.
We will further explore the different choices of measures in Section~\ref{sec:Some_Measures}.

The Cycle Walk is a Markov chain with state space $\mathcal{F}_d$. In its more basic form, it will be a mixture of two other Markov chains: the 1-Tree Cycle Walk $Q_\treeOne$ and the 2-Tree Cycle Walk $Q_\treeTwo$. Though their constructions are very similar, they play very different roles in the Cycle Walk.
Once we define these Markov chains, we will define the $\kappa$-Cycle Walk by $Q_{\text{Cycle},\kappa}= \kappa Q_{\treeOne}+ (1-\kappa) Q_{\treeTwo}$ for $\kappa \in [0,1)$. Often we will fix $\kappa$ and suppress it from the notation, writing simply $Q_{\text{Cycle}}$. Though we begin by defining the simple Cycle Walk, the main object of interest for us will be the Metropolized Cycle Walk, which we will denote by $P_{\text{Cycle},\kappa}$ and will define at the end of the section. As before, Metropolization will be used to make the chain have the specified target measure $\nu$ on  $\mathcal{F}_d(G)$ as its stationary measure.

The 1-Tree and 2-Tree Cycle Walks are built on the common idea of adding a cycle to a tree or pair of trees and then removing edges to return to a forest.  We begin by defining these two building blocks.

\subsection{The 1-Tree Cycle Walk} \label{sec:1Tree} We will assume that we are given a collection of edge weights $\alpha\colon E \rightarrow (0,\infty)$ with which to construct the 1-Tree Cycle Walk. The 1-Tree Cycle Walk on a forest $\tau = \{t_1, \dots,t_d\} \in \mathcal{F}_d$ begins by choosing a random edge that is (i) contained within an induced subgraph of one of the trees, and (ii) does not belong to one of the trees, i.e. an edge $e=(u,v)$ such that there is some tree, $t_i$ with $u\in V(t_i)$, $v\in V(t_i)$ and $(u,v)\notin E(t_i)$. Unless otherwise stated, we will choose the edge uniformly at random. 

\begin{figure}[htbp]
    \centering 

    \begin{subfigure}[t]{0.39\textwidth}
        \includegraphics[width=\linewidth]{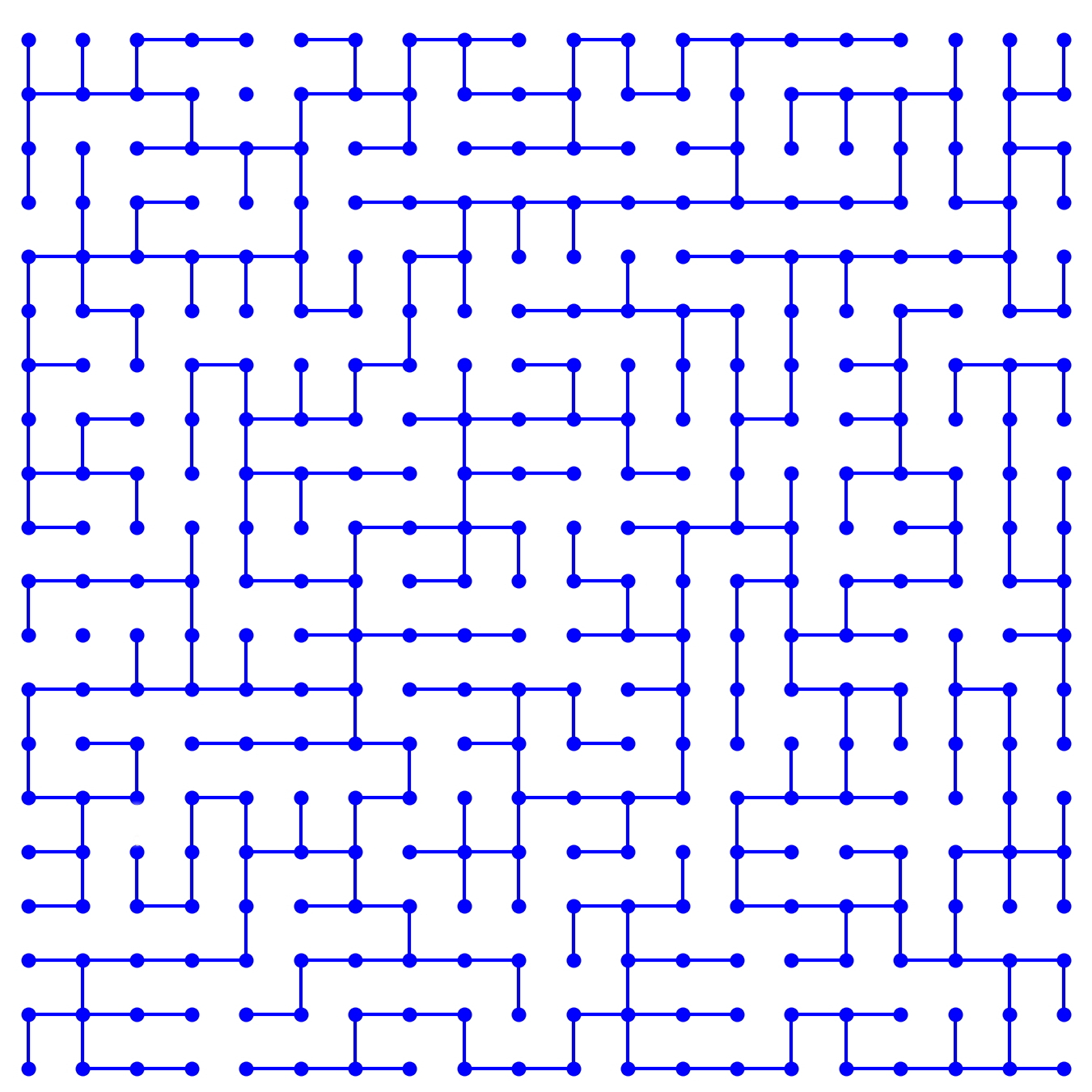}
        \caption{Start with a spanning tree.}
    \end{subfigure}
    \hfill 
    \begin{subfigure}[t]{0.39\textwidth}
        \includegraphics[width=\linewidth]{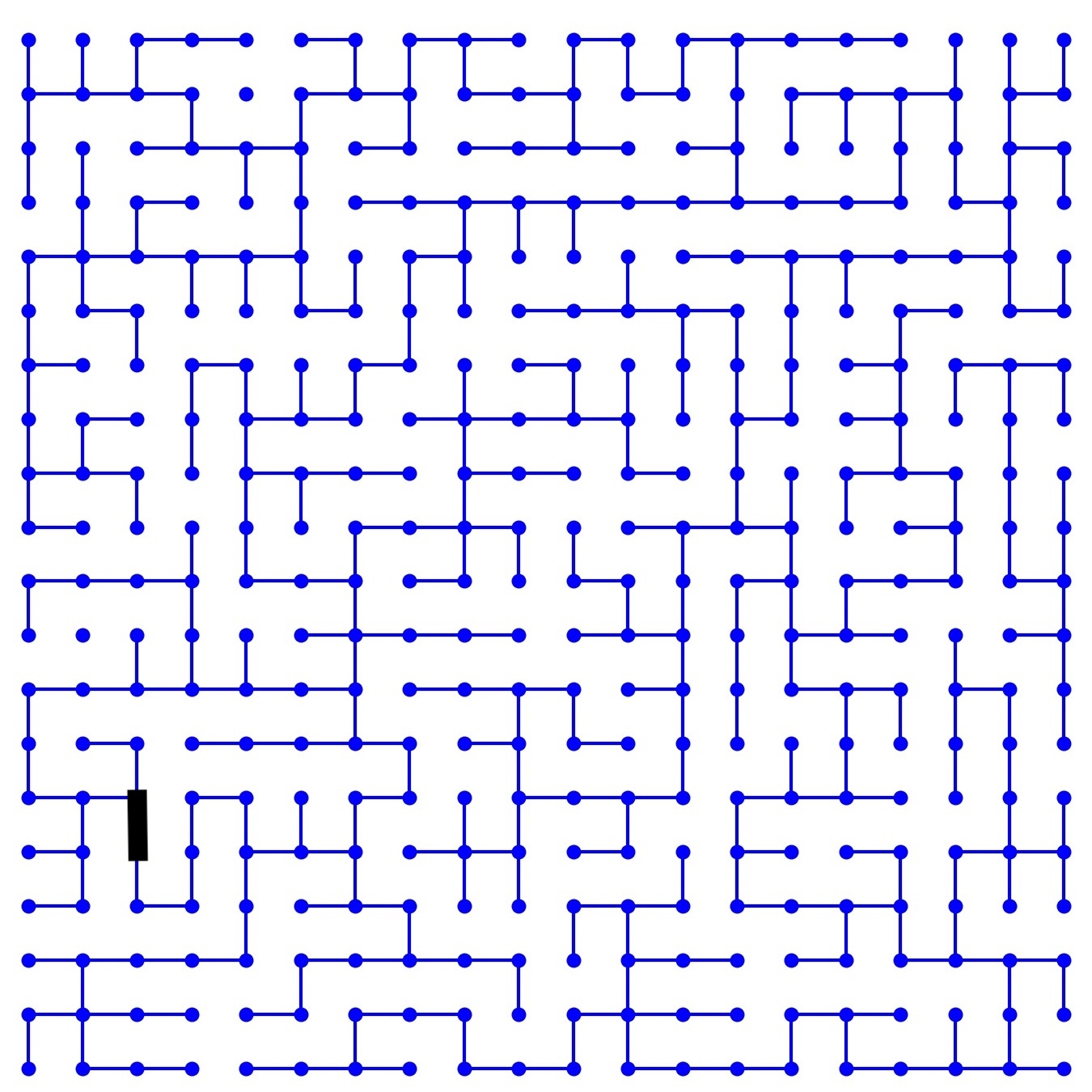}
        \caption{Add a missing edge (black) uniformly at random to the existing tree (blue).}
    \end{subfigure}

    \vspace{.5em}

    \begin{subfigure}[t]{0.39\textwidth}
        \includegraphics[width=\linewidth]{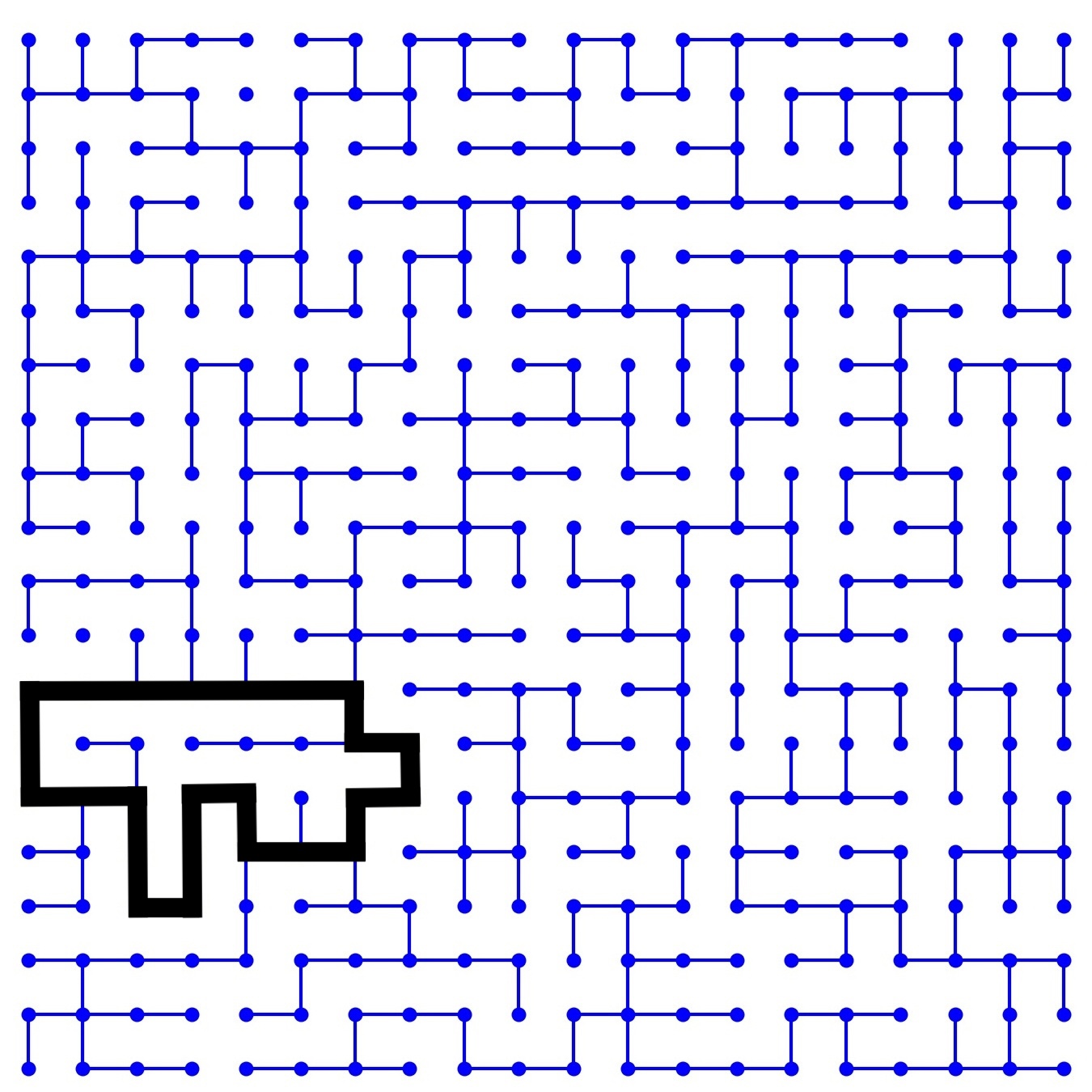}
        \caption{Find the cycle  (black) the new edge induces.}
    \end{subfigure}
    \hfill 
    \begin{subfigure}[t]{0.39\textwidth}
        \includegraphics[width=\linewidth]{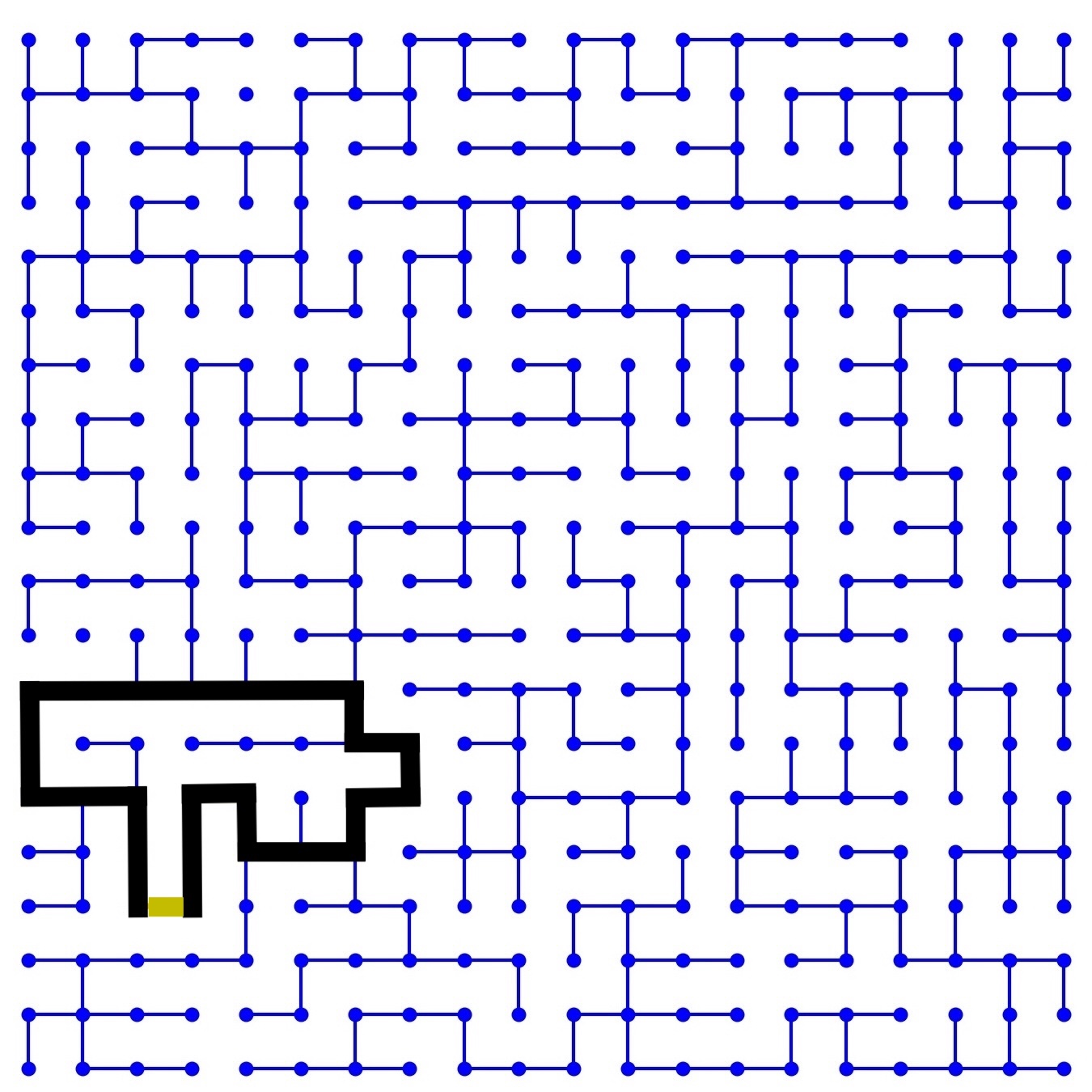}
        \caption{Choose uniformly at random an edge (yellow) from the cycle  (black) to remove.}
    \end{subfigure}

    \vspace{.5em}
    
   \begin{subfigure}[t]{0.39\textwidth}
        \includegraphics[width=\linewidth]{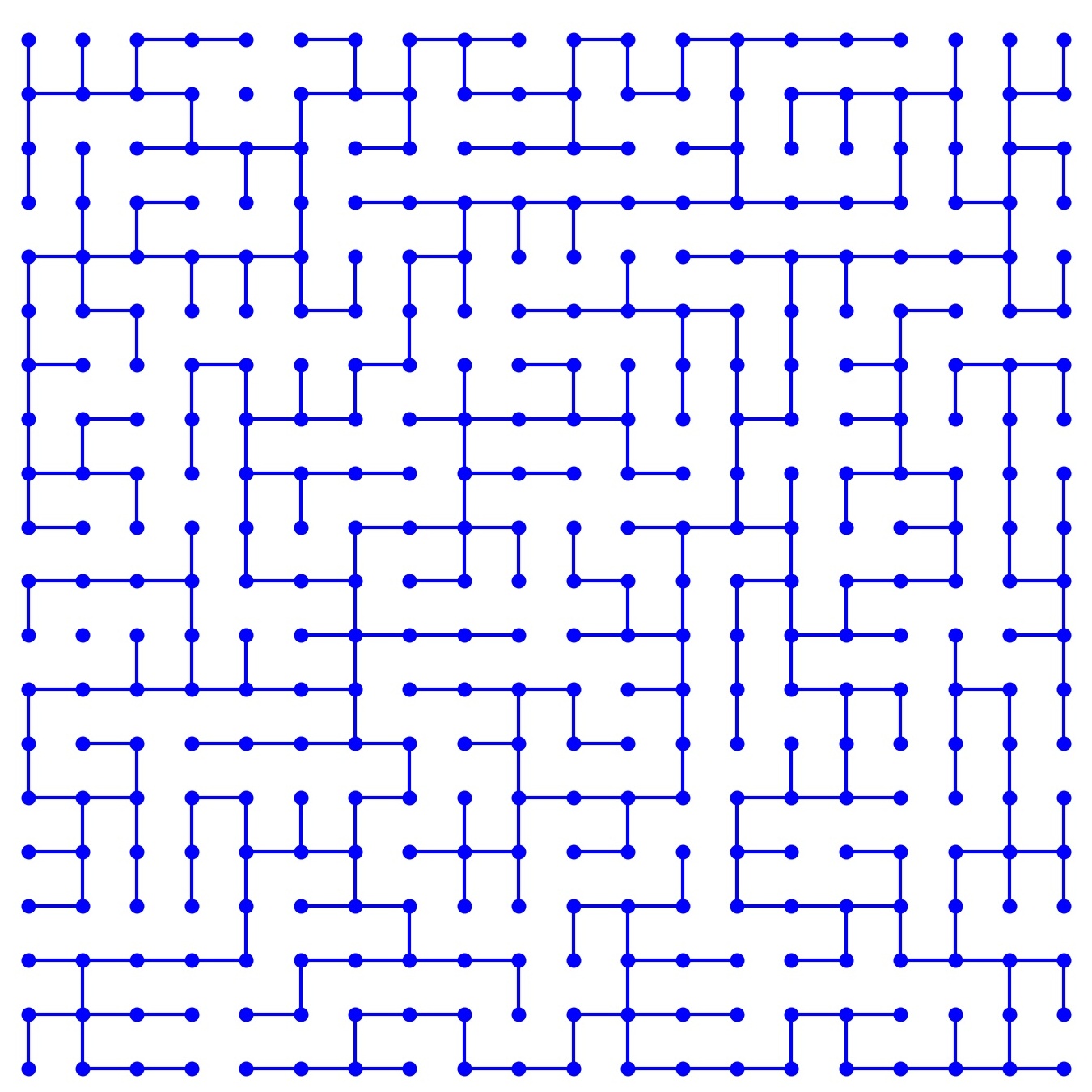}
        \caption{The resulting new spanning tree obtained by removing the (red) edge chosen in the previous step.}
      \end{subfigure}
      \caption{Illustration of one step of the 1-Tree Cycle Walk}
      \label{fig:1treecycle}
\end{figure}

The second step is to choose a random edge $e \in E(G_i) \cap E_i^c$; that is to say, an edge in $G_i$ but not the tree $t_i$. There are a number of ways to choose this edge; unless otherwise stated, we will choose the edge uniformly at random from $E(G_i) \cap E_i^c$.   
Adding this edge to $t_i$ creates a unique cycle in $t_i \cup e$, which we will denote $C$. We then choose a random edge from $C$ to remove, which will create a new tree $t_i'$ out of $t_i \cup e$. We will choose a new edge $e'$ randomly from the cycle $C$ with a probability proportional to $1/\alpha(e')$. The new forest $\tau'$ obtained from $\tau$ by replacing $t_i$ with $t_i'$ is one step of the kernel $Q_\treeOne$. \Cref{fig:1treecycle} shows the steps of the 1-Tree Cycle Walk.

At times, we will refer to the 1-Tree Cycle Walk as the internal Cycle Walk. This is because the 1-Tree Cycle Walk only changes a tree in a forest, but not the partition that the forest induces. This implies that $Q_\treeOne(\tau, \tau_{\xi_\tau})=1 $. If $\xi=\{D_1,\cdots,D_d\} \in \mathcal{P}_d$, it is reasonable to ask what is the invariant measure of $Q_\treeOne$ acting on $\mathcal{F}_d^\xi\eqdef  \{ \tau \in \mathcal{F}_d: \xi_\tau=\xi\}$. Fixing a weight $\alpha$ used in defining the 1-Tree Cycle Walk, from \cite{russo2018linking, AnariVinzantVuong2021}, 
one can see that the 1-Tree Cycle Walk is reversible with respect to the product probability measure $m^{(\alpha)}_\xi$ defined for $\tau=\{t_1,\cdots,t_d\} \in \mathcal{F}_d^\xi$ by 
\begin{equation}\label{eq:mAlpha}
  m^{(\alpha)}_\xi(\tau) = m^{(\alpha)}_\xi(t_1,\cdots,t_d) = \prod_{i=1}^d m_{D_i}^{(\alpha)}(t_i) \quad\text{where}\quad m_{D_i}^{(\alpha)}(t_i)= \frac{\Tree_\alpha(t_i)}{\Tree_\alpha(D_i)}
\end{equation}
and
\begin{equation}
  \label{eq:TreeAlpha}
  \Tree_\alpha(D_i)=\sum_{t \in  {\tau_D}_i}\Tree_\alpha(t) \quad\text{and}\quad \Tree_\alpha(t)= \prod_{ e \in E(t)} \alpha(e), 
\end{equation}
which is to say that the probability of seeing a forest on a particular partition, $\xi$, is proportional to a product of the edge weights within the forest (see \eqref{eq:TreeAlpha}).
We will also extend the definitions of $\Tree_\alpha$ to be defined on both partitions and forests as
\begin{align}
\Tree_\alpha(\xi) = \prod_{i=1}^{d} \Tree_\alpha(D_i), \quad\text{and}\quad 
\Tree_\alpha(\tau) = \prod_{i=1}^{d} \Tree_\alpha(t_i),
\label{eq:treealphaOnStates}
\end{align}
which gives the equivalent expression $m^{(\alpha)}_\xi(\tau)=\frac{\Tree_\alpha(\tau)}{\Tree_\alpha(\xi)}$ for $\tau \in \mathcal{F}^\xi_d$.

For any partition $\xi \in \mathcal{P}_d$ and forests that induce the partition, $\tau, \tau' \in \mathcal{F}_d^\xi$, a direct computation shows that the chain satisfies detailed balance, i.e., 
\begin{align*}
m^{(\alpha)}_\xi(\tau) Q_\treeOne(\tau,\tau') = m^{(\alpha)}_\xi(\tau') Q_\treeOne(\tau',\tau).
\end{align*}
This implies that  $m^{(\alpha)}_\xi$ is an invariant measure of $Q_\treeOne$ with weights $\alpha$.
When $\alpha$ is a constant, we will denote $\alpha$ as $1$ and write $m^{(1)}_\xi$. Notice that, conditioned on a fixed partition $\xi$,  $m^{(1)}_\xi(\tau | \xi)$ is the uniform measure on $\mathcal{F}_d^\xi$ with $m^{(1)}_\xi(\tau)=1/\Tree(\xi)$ for all $\tau \in \mathcal{F}_d^\xi$.  This implies $Q_\treeOne$ with constant $\alpha$ simply exchanges equiprobable trees within fixed districts. 
$\nu_\gamma(\tau)=\nu_\gamma(\tau')$ for all $\tau, \tau' \in \mathcal{F}_d^\xi$ (i.e., $\xi_\tau=\xi_{\tau'}$) for any $\xi \in \mathcal{P}_d$.

We can also generalize the measure on the forests, $\mathcal{F}_d(G)$, by letting $\nu(\tau) = \pi(\xi_\tau) M_\xi(\tau)$ for probability measure $\pi$ on $\mathcal{P}_d$ and some collection of probability measures $M_\xi$ on $\mathcal{F}_d^\xi$ for all $\xi \in \mathcal{P}_d$. Then for any $\xi \in \mathcal{P}_d(G)$, $Q_\treeOne$ will leave unchanged the projection of $\nu$ onto $\mathcal{P}_d(G)$, i.e., it will fix the districts, but alter the trees that define them. However, the $Q_\treeOne$ walk will no longer be invariant with respect to more generalized measures.  

\subsection{The 2-Tree Cycle Walk}\label{sec:2Tree} 
\begin{sloppypar} The 2-Tree Cycle Walk on a forest $\tau = \{t_1, \dots,t_d\} \in \mathcal{F}_d$ begins by choosing two neighboring trees, $t_i=(V_i,E_i)$ and  $t_k=(V_k,E_k)$, at random according to some distribution.  Unless otherwise stated, we will choose the pair $t_i$ and $t_k$ uniformly at random among all neighboring pairs of trees in $\tau$. Next we randomly pick two edges $e_1, e_2 \in (\partial t_i) \cap  (\partial t_k)$. Unless otherwise stated, we will choose the pair uniformly at random among all such pairs. Adding the edges $e_1$ and $e_2$ to $t_i$ and $t_k$ creates a unique cycle $C$ in the graph $G'=t_i \cup t_k\cup e_1\cup e_2$.  We now choose two edges $e_1'$ and $e_2'$ to remove from $G'$. Doing so necessarily splits $G'$ into two new trees $t_i'$ and $t_k'$. 
\end{sloppypar}

\begin{figure}[htbp]
    \centering 

    \begin{subfigure}[t]{0.39\textwidth}
        \includegraphics[width=\linewidth]{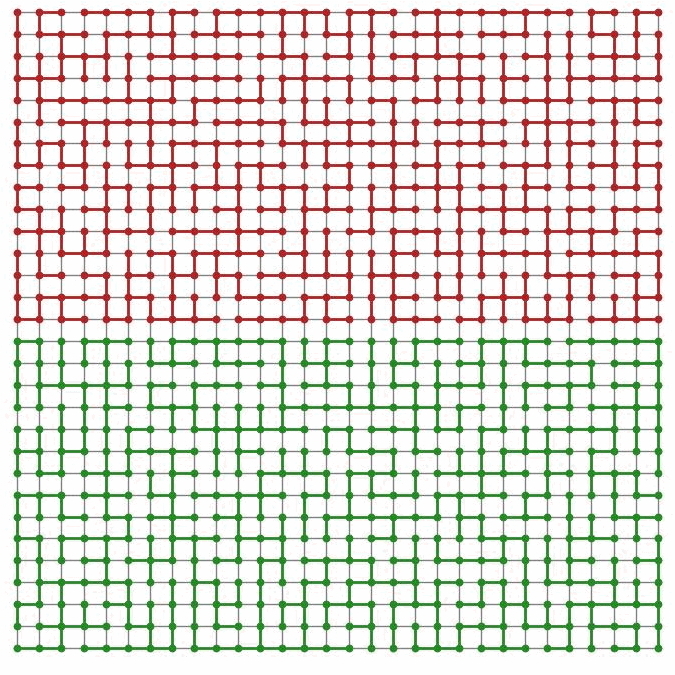}
        \caption{A spanning forest with two balanced trees (red and blue).}
    \end{subfigure}
    \hfill 
    \begin{subfigure}[t]{0.39\textwidth}
        \includegraphics[width=\linewidth]{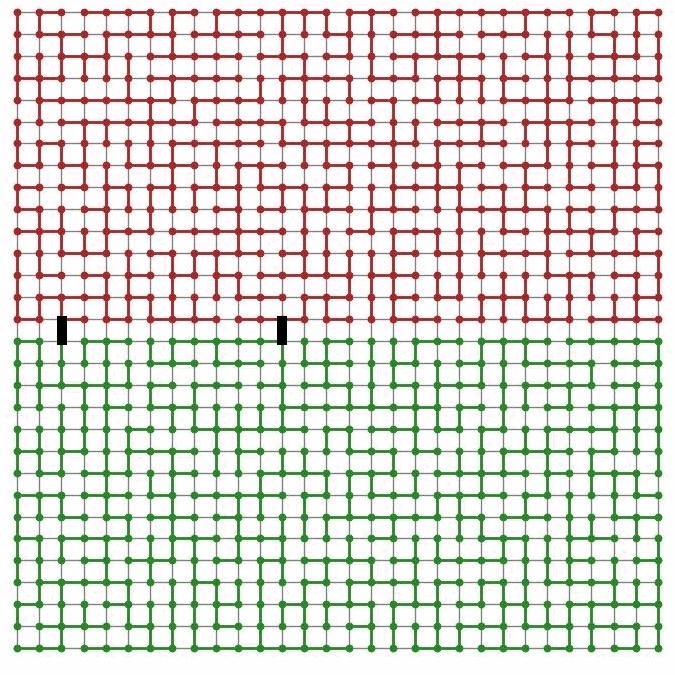}
        \caption{Add uniformly at random two edges that connect the two trees (thick black).}
    \end{subfigure}

    \vspace{.5em}

    \begin{subfigure}[t]{0.39\textwidth}
        \includegraphics[width=\linewidth]{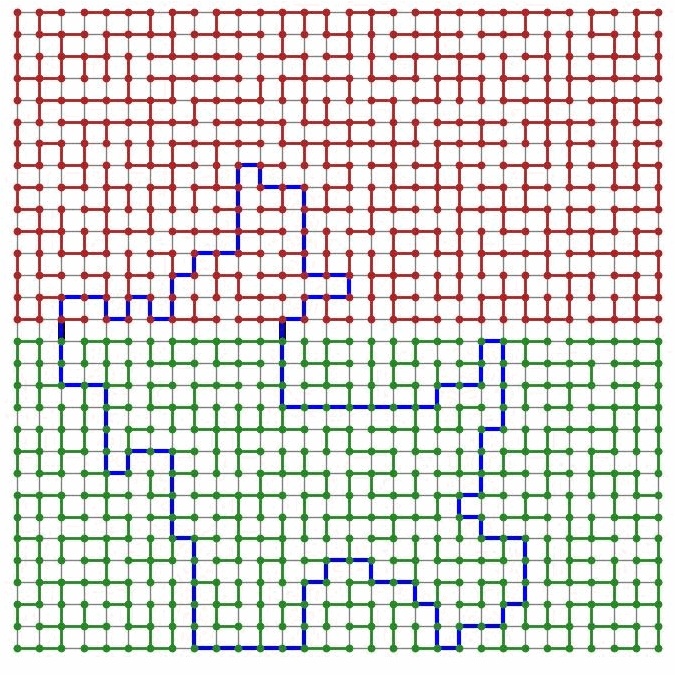}
        \caption{Find the unique cycle (dark blue) with the two added edges (thick black).}
    \end{subfigure}
    \hfill 
    \begin{subfigure}[t]{0.39\textwidth}
        \includegraphics[width=\linewidth]{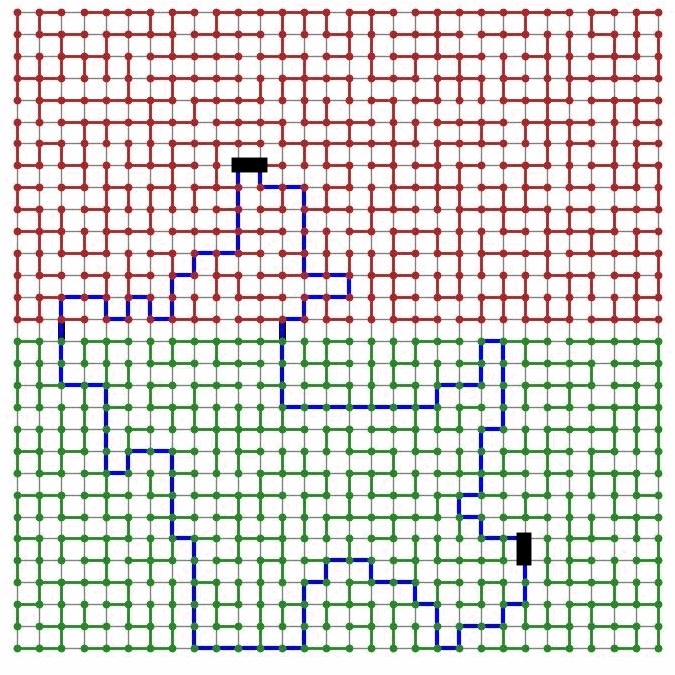}
        \caption{Pick two edges (light blue) to remove from the cycle among all those that leave the resulting two trees balanced. }
    \end{subfigure}

     \vspace{.5em}
    
    \begin{subfigure}[t]{0.39\textwidth}
        \includegraphics[width=\linewidth]{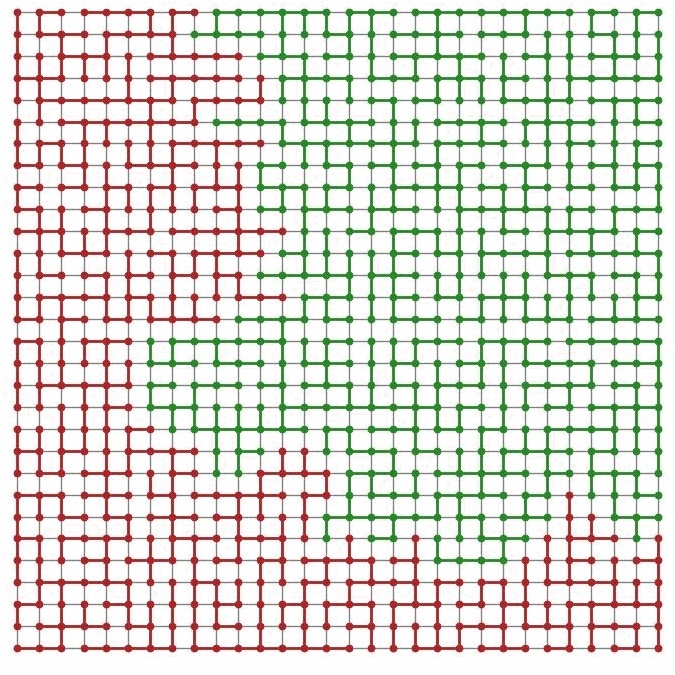}
        \caption{The new resulting balanced forest with two trees.}
      \end{subfigure}
      \caption{Illustration of the 2-Tree Cycle Walk.}
      \label{fig:2treecycle}
\end{figure}

In selecting edges to remove from the cycle, we will assume that we are given a collection of edge weights $\beta\colon E \rightarrow (0,\infty)$ with which to construct the 2-Tree Cycle Walk. 
We choose the edges $e_1'$ and $e_2'$ to remove by selecting at random with a probability proportional to $1/(\beta(e_1')\beta(e_2'))$  among all pairs of edges in the cycle $C$ that, when removed, create two trees that respect the specified population balance bounds. The new forest $\tau'$ obtained by replacing $t_i$ and $t_k$ in $\tau$ with $t_i'$ and $t_k'$ represents one step of the Markov chain  $Q_\treeTwo$. Notice that   $Q_\treeTwo$ generally will not preserve the measure $\nu_\gamma$ for any $\gamma$. \Cref{fig:2treecycle} shows one step of this chain.

Both the $Q_\treeOne$ and $Q_\treeTwo$ walks are closely related to the Up-Down Walk on spanning forests \cite{russo2018linking, anari2021logconcavepolynomialsivapproximate}. In the Up-Down Walk, two trees in the forest are joined by adding a single edge. If initially the forest was an element of $\mathcal{F}_d$, then after the joining move, the forest is an element of $\mathcal{F}_{d-1}$. One can choose one of the trees in the forest to split in two by removing a single edge. The resulting forest will be in  $\mathcal{F}_{d}$. The term ``Up-Down'' refers to the fact that one moves from $\mathcal{F}_{d}$ up to $\mathcal{F}_{d-1}$ adding an edge, the ``up'' step, and then back to $\mathcal{F}_{d}$ by removing an edge, the ``down'' step.

When sampling cycles across adjacent districts with the 2-Tree Cycle Walk, the cycle lengths and possible ways to cut the cycle are capable of making both large and small structural changes. Thus, we preserve RECOM's ability to make large-scale changes while also preserving the ability to make the small localized changes that occur in the Flip Walk/Single Node Flip algorithm, all while maintaining population balance. Like the Swendsen–Wang algorithm \cite{SwendsenWang,fifield2020automated}, the range of allowable changes allows the algorithm to adapt to particular measures and graph structures. The Cycle Walk can thus make locally adaptive moves while also allowing global moves to accelerate escaping energetic barriers. For these reasons, we expect the Metropolized Cycle Walk to be able to sample from a wider range of measures than some other methods. This is supported by our numerical experiments in \Cref{sec:ratesGammaNC}  and \Cref{sec:ConvergenceRegularLattices}, which show that Cycle Walk mixes well for values of $\gamma$ much closer to one when compared with RECOM. In \Cref{sec:structureCycleWalk}, we see that Cycle Walk proposes moves at a variety of scales that may allow more proposed moves to be accepted across a range of $\gamma$ and other modifications to the score function.

\subsection{\texorpdfstring{$k$}{k}-Tree Cycle Walk} We now describe a generalization to a $k$-Tree Cycle Walk. For clarity, we describe one among many possible generalizations. In particular, we will not include edge weights in this version, though it is straightforward to do so.

Using some method, choose a simple loop $t_{j_1},\dots, t_{j_k}$ among the trees in a forest $\tau=\{t_1,\dots,t_d\}$. That is, $t_{j_a} \neq t_{j_b}$ if $a \neq b$ and for all $a$, $t_{j_a}$ and $t_{j_{a+1}}$ are adjacent with the convention that $t_{j_1}=t_{j_{k+1}}$. We now pick exactly $k$ edges not in our trees to add so that the $k$ selected trees are now all connected to their neighbor in the loop of trees. By construction, the resulting connected subgraph has a unique cycle. We then choose $k$ edges to remove from this cycle uniformly among all $k$ edges whose removal produces a forest $\tau' \in \mathcal{F}_{d}$ that has trees with populations within the prescribed tolerance.
While we think that the $k$-Tree Cycle Walk provides an interesting generalization that may improve mixing and the ability to prove mixing bounds, we do not explore it here because we would need to look for all $k$-tuples of edges to remove to achieve balance, and this set may grow exponentially with $k$. Thus, we only look for pairs or singles with the 2-Tree or 1-Tree Cycle Walks. We do discuss an alternative method for including $k$-Cycles in Section~\ref{sec:updown}, but do not implement it in the current work. It is the analogy of the $k$ district RECOM that has been used in some settings. 

\section{The Metropolized Cycle Walk}

We now discuss Metropolizing the Cycle Walk to preserve a specified measure on $\mathcal{F}_d(G)$. We begin by defining a more general measure on $\mathcal{F}_d(G)$ that agrees with $\pi$ when projected down to the space of partitions $\mathcal{P}_d(V)$. This will be a general case of some of the constructions already made in \Cref{sec:1Tree}. Recall that in \Cref{sec:1Tree}, for any $\xi \in \mathcal{P}_d$ we defined in \eqref{eq:Fd} $\mathcal{F}_d^\xi$ to be the space of all spanning forests that imply the same partition $\xi$. Namely if $\tau \in \mathcal{F}_d^\xi$, then $\xi_\tau=\xi$.

Given any subset $U \subset V$ of vertices that induces a connected subgraph $H$ in $G$, we define a measure $\mu_U$ on the space of spanning trees of $H$. Given a partition $\xi=\{D_1,\dots,D_d\} \in \mathcal{P}_d$, we define the measure $\mu_\xi$ on $\mathcal{F}_d^\xi$ by
\begin{align*}
  \mu_\xi(\tau)=\prod_{i=1}^d \mu_{D_i}(t_i) \quad\text{for}\quad\tau=\{t_1,\dots,t_d\} \in \mathcal{F}_d^\xi,
\end{align*}
e.g. the measure $m^{(\alpha)}_\xi$ defined above.
Then given a measure $\pi$ on $\mathcal{P}_d$, we define the probability measure
\begin{align*}
  \nu(\tau) \propto \pi(\xi_\tau) \mu_{\xi_\tau}(\tau).
\end{align*}
If we require that each of the  $\mu_{U}$ is normalized to be a probability measure, which can always be done by incorporating any normalization into $\pi$, then the projection of $\nu$ from $\mathcal{F}_d$ to $\mathcal{P}_d$ will coincide with $\pi$.

We will now use the Metropolis-Hastings algorithm with proposal kernels $Q_{\treeOne}$ and $Q_{\treeTwo}$ to define Markov chains with kernels $P_{\treeOne}$ and $P_{\treeTwo}$, respectively.  By design  $P_{\treeOne}$ and $P_{\treeTwo}$ will both have $\nu$ as a stationary measure.   We will then define $P_{\text{Cycle},\kappa}= \kappa P_{\treeOne}+ (1-\kappa) P_{\treeTwo}$ for $\kappa \in [0,1)$.

We recall that given a proposal Markov transition kernel $Q$ and a target measure $\nu$ on a discrete state space $\X$, the Metropolis-Hastings transition kernel is defined by the following procedure. If the current state is $X_n \in \X$ then: 
\begin{enumerate}
\item we draw the proposal $X_n'$ according to the measure $Q(X_n,\ccdot)$.
\item we accept the proposal $X_n'$ with probability $ a(X_n,X_n')$ where
  \begin{align*}
   a(x,x')= 1\wedge \frac{\nu(x')Q(x',x)}{\nu(x)Q(x,x')}.
  \end{align*}
\item if the proposal is accepted we set $X_{n+1}=X_n'$ and if the proposal is not accepted we set  $X_{n+1}=X_n$.
\end{enumerate}
\subsubsection*{Metropolized 1-Tree Cycle Walk} The Metropolized 1-Tree Cycle Walk's Markov Transition Kernel will be denoted by  $P_{\treeOne}$. It will be constructed using the Metropolis-Hastings algorithm with proposal kernel  $Q_{\treeOne}$  and
target measure $\nu$.  In the case when $\nu$ is of the form given in \eqref{eq:nu_0_1}, no Metropolization is needed and this section can be ignored.
Given any two $\tau=\{t_1,\dots,t_d\}, \tau'=\{t_1',\dots,t_d'\} \in \mathcal{F}_d$, we need only define the acceptance probability $a_\treeOne(\tau,\tau')$ when $Q_\treeOne (\tau,\tau')>0$. For completeness we will define $a _\treeOne(\tau,\tau')=0$ when $Q_\treeOne(\tau,\tau')=0$. When $Q_\treeOne(\tau,\tau')>0$, $\tau$ and $\tau'$ must only differ by a single tree. Let $t$ be the tree in $\tau$ that is not in $\tau'$ and $t'$ the tree in $\tau'$ but not in $\tau$. By the construction of   $Q_\treeOne$, $t$ and $t'$ have the same vertices $U \subset V$. Since $\xi_\tau=\xi_{\tau'}$, $U$ is just the  element  of this partition that contains $t$ (and necessarily $t'$). Furthermore, there exists an edge $e \in E(t)$ and an edge $e' \in E(t')$ so that $t = (t \cap t')\cup e$ and $t' = ( t \cap t')\cup e'$. 
\begin{align*}
  a_\treeOne(\tau,\tau')&= 1\wedge \frac{\nu(\tau') Q_\treeOne (\tau',\tau)}{\nu(\tau) Q_\treeOne (\tau,\tau')}
  =1 \wedge \frac{\alpha(e)}{\alpha(e')}\frac{\mu_{U}( t')}{\mu_{U}( t)}.
\end{align*}
Utilizing $m^{(\alpha)}_\xi$ for the measure on the forest conditioned on the partition, we recover the special case from \cite{russo2018linking,AnariVinzantVuong2021} where the acceptance ratio is always 1 (see \Cref{sec: Implemented} below for more detail).

\subsubsection*{Metropolized 2-Tree Cycle Walk}  The Metropolized 2-Tree Cycle Walk's Markov Transition Kernel will be denoted by  $P_{\treeTwo}$. Again, it will be constructed using the Metropolis-Hastings algorithm with  proposal kernel  $Q_{\treeTwo}$  and
target measure $\nu$. Given any two $\tau, \tau' \in \mathcal{F}_d$, as before we need only define the acceptance probability $a_{\treeTwo}(\tau,\tau')$ when $Q_{\treeTwo}(\tau,\tau')>0$. For completeness we will define $a_{\treeTwo}(\tau,\tau')=0$ when $Q_{\treeTwo}(\tau,\tau')=0$. When $Q_{\treeTwo}(\tau,\tau')>0$, $\tau$ and $\tau'$ must have exactly $d-2$ trees in common.  Let $t_a$ and $t_b$ be the two trees in $\tau$ that are not in $\tau'$ and  $t_a'$ and $t_b'$ be the two trees in $\tau'$ that are not in $\tau$. By the construction of $Q_{\treeTwo}$, $t_a$ and $t_b$  must be adjacent and  $t_a'$ and $t_b'$  must be adjacent.  Furthermore $G(t_a\cup t_b)= G( t_a'\cup t_b')$ and there exist $e_a, e_b \in t_a\cup t_b$ and $e_a', e_b' \in t_a'\cup  t_b'$ so that $t_a\cup t_b \cup e_a' \cup e_b' = t_a' \cup t_b' \cup e_a \cup e_b$. Let $\text{adj}(\tau)$ be the pairs of trees in $\tau$ which are adjacent on the graph and $|\text{adj}(\tau)|$ be the number of pairs of trees. Lastly, recall that $(\partial  t_a) \cap (\partial t_b)$ are the edges that make up the boundary between the trees $t_a$ and $t_b$. 
Hence if
\begin{align*}
  B= | (\partial t_a) \cap (\partial t_b)| \qquad \text{and}\quad B'= | (\partial  t_a') \cap (\partial t_b')|,
\end{align*}
then $B(B-1)$ is twice the number of pairs of border edges between $t_a$ and $t_b$ and  $B'(B'-1)$ twice the number of  pairs between  $t_a'$ and $t_b'$.  Combining everything, we have
\begin{align}
\nonumber
  a_{\treeTwo}(\tau,\tau')&= 1 \wedge \frac{\nu(\tau')Q _{\treeTwo} (\tau',\tau)}{\nu(\tau) Q _{\treeTwo} (\tau,\tau')} \\
                                 &=
  1 \wedge \frac{|\text{adj}(\tau)|}{|\text{adj}(\tau')|}\frac{B(B-1)}{ B'(B'-1) }\frac{ \beta(e_a)\beta(e_b) }{\beta(e_a')\beta(e_b')}\frac{\mu_{U_a'}(t'_a) \mu_{U_b'} (t_b')}{ \mu_{U_a}(t_a) \mu_{U_b}(t_b) }
   e^{J(\xi_{\tau}) -J(\xi_{\tau'})}.
   \label{eq:accept2treegeneral}
\end{align}
where $U_a=V(t_a)$,  $U_b=V(t_b)$, $U_a'=V(t_a')$ and  $U_b'=V(t_b')$.

\section{Some Measures on Spanning Forests}\label{sec:Some_Measures}
In this work, we will primarily investigate the class of measures defined in \eqref{eq:nudefinition}. We will consider $J=0$ and $J=w J_{\Compact}$, for some weight $w\in \R$, as explained below in \eqref{eq:compactscore}. When lifting the measure $\pi$ from the space of partitions to the space of spanning forest, one can choose to weight each element in $\tau_\xi$, for a given partition $\xi$, with the same weight or introduce some internal structure in side of $\tau_\xi$. One convenient way to do this is to introduce a probability measure $m_\xi$ on  $\tau_\xi$. Then $ \tau \mapsto \pi(\tau_\xi) m_{\tau_\xi}(\tau)$ is a measure the space of spanning forest that induces the same distribution on partitions as $\pi$. In the context of   $\nu_\gamma$, from \eqref{eq:nudefinition}, the conditional probability measure  $m^{(\alpha)}_\xi$  described in \eqref{eq:mAlpha}  plays this role.  Alternatively, one can recast the defining of the measure on spanning forest by introducing a score function. We will see that this second approach does not necessarily produce a measure whose projection to the space of partitions agrees with $\pi$.   

If we let $J_{\Tree}(\xi) = \log \Tree(\xi)$, then
\begin{align*}
\nu_\gamma(\tau) \propto\frac{\pi(\xi_\tau)}{\Tree(\xi_\tau)^\gamma} = \frac{\pi(\xi_\tau)}{\Tree(\xi_\tau)^{\gamma-1}} m^{(1)}_{\xi_\tau}(\tau) \propto e^{-\gamma J_\Tree(\xi_\tau) - J(\xi_\tau)},
\end{align*}
where $m^{(1)}_\xi(\tau)$ is the conditional measure described in \eqref{eq:mAlpha} with constant weights one. 
The first expression may be thought of as some underlying measure $\pi$ being modified by a base measure $\Tree^{-\gamma}$.
The second expression is the probability of a partition multiplied by the probability of a tree conditioned on the partition. The final exponential expression considers the tree counts to be part of the score function that can be used as a policy consideration; several works explicitly use the spanning forest count as a novel definition of compactness, (e.g. \cite{deford2019recombination,mccartan2020sequential}).

The measures $\nu_\gamma$ are invariant measures for the 1-Tree Cycle Walk for any $\gamma$; if, however, we want to utilize a collection of weights $\alpha$ in the definition of the 1-Tree or 2-Tree Cycle Walks then $\nu_\gamma$ will not be an invariant measure for the 1-Tree Cycle Walk, though the projection of $\nu_\gamma Q_{\treeOne}$ and $\nu_\gamma$ onto $\mathcal{P}_d$ will coincide as already discussed at the end of Section \ref{sec:1Tree} where the 1-Tree Cycle Walk was introduced.  However, if we define 
\begin{align*}
  \nu_\gamma^{(\alpha)}(\tau) \propto \frac{\pi(\xi_\tau)}{\Tree_\alpha(\xi_\tau)^\gamma}
\end{align*}
then $\nu_\gamma^{(\alpha)}$ is invariant for the  1-Tree Cycle Walk with weights $\alpha$.

Just like the unweighted case, there are several equivalent representations of the weighted measure, given as 
\begin{align}
\nu_\gamma^{(\alpha)}(\tau) = \frac{\pi(\xi_\tau)}{\Tree_\alpha(\xi_\tau)^\gamma} = \frac{\pi(\xi_\tau)}{\Tree_\alpha(\xi_\tau)^{\gamma-1}} m^{(\alpha)}_{\xi_\tau}(\tau) \propto e^{-\gamma J_{\Tree}^{(\alpha)}(\xi_\tau) - J(\xi_\tau)},
\label{eq:equivnualphas}
\end{align}
where $m^{(\alpha)}_{\xi_\tau}$ was the probability measure defined in \eqref{eq:mAlpha}, 
and $J_{\Tree}^{(\alpha)}(\xi)= \log \Tree_\alpha(\xi)$ where $\Tree_\alpha$ was defined in \eqref{eq:treealphaOnStates}.

\section{The Need to Move Beyond Metropolized Forest Recombination}\label{sec:BeyondRECOM}

One of the strengths of the RECOM-based procedure is also a large limitation. Because the RECOM chain replaces the entire spanning tree in the merged region with an independent new spanning tree, which is then split, the process can appear to mix very fast. It can be seen as a kind of conditionally ``independent'' sampler which replaces a large part of the state space with a new independent chunk. If one then cuts the merged region in two by removing an edge, one will always end up with a boundary that has the structure obtained by cutting a randomly chosen spanning tree in two. One can influence, to some degree, the structure of the randomly chosen spanning tree on the merged region. There are at least two methods for efficiently drawing random trees.  Wilson's algorithm \cite{wilsonGeneratingRandomSpanning1996} can efficiently draw Uniform Spanning Trees (UST) while Kruskal's algorithm, coupled with random edge weights \cite{Kruskal56,Cormen} generates independent samples from the minimum spanning tree (MST) distribution \cite{babson2024modelsrandomspanningtrees}. Each of these distributions has different properties. Some of these properties are desirable. As emphasized in \cite{deford2019recombination}, Uniform Spanning Trees tend to have nice compactness properties. In both UST and MST, one can introduce weights \cite{babson2024modelsrandomspanningtrees,Clelland} between edges to encourage particular vertices to be connected or not in the spanning tree. One can introduce a probability measure on the edge removed to cut the tree in two, so that certain structures are encouraged. 

However, as seen in numerical examples \cite{spectralAnalysis}, when the natural structures of the target measure $\pi$ from \eqref{eq:pi_def} (or equivalently $\nu_1$ from \eqref{eq:nu_0_1} or \eqref{eq:nudefinition}) are not aligned with the proposal measure, the acceptance rate can plummet to zero as the problem size increases. This is also borne out in the numbers in Figure~\ref{fig:convergenceWGammaStudy} where we see that Metropolized Forest Recombination has difficulty converging at much smaller $\gamma$ than Cycle Walk.  One solution is just to take what the proposal chain (or some weighted version) gives you and not try to correct it by metropolization. The difficulty with this is that the ability to tune toward measures described directly in terms of policy goals can be limited \cite{autry2020multiscale,autry2021metropolized,zhao2022mathematically}. 
Different policy choices, for example, in the choice of compactness measure, have impacts on the redistricting space. For example, in \cite{jonasBlogPost}, it was shown that using the spanning tree counts (as in RECOM) as the measure of compactness can create differences in how rural-urban interfaces are treated relative to urban-urban or rural-rural interfaces as compared with a Polsby-Popper compactness score. The ability to tune and test a variety of measures allows us to align these measures with stated policy goals. 

By making a wide range of moves ranging from big to small, the Cycle Walk chain can sample from a broader class of measures via Metropolization. 
This is further described by the results in Section~\ref{sec:structureCycleWalk}.

\section{The Particular Flavor of Cycle Walk and Target Measures Implemented}\label{sec: Implemented}
We now present the particular algorithm that we implemented for our numerical experiments. While the version described here fits
completely in the framework from \Cref{sec:cycleWalk}, it contains a number of simplifications that exist in many applications and leads to additional interpretability and properties. We will only be interested in target measures $\nu$ on  $\mathcal{F}_d$ that come from a lift measure $\pi$ on  $\mathcal{P}_d$. In particular if $\pi$ is given by \eqref{eq:pi_def}, then we will be interested in sampling $\nu_\gamma$ of the form given in \eqref{eq:nudefinition} for some $\gamma \in \R$ (though typically $\gamma \in [0,1]$) but with $\Tree$ replaced with $\Tree_\alpha$ from \eqref{eq:TreeAlpha} for the weights $\alpha$ used to define the 1-Tree Cycle Walk. That is to say,
\begin{align*}
\nu_\gamma^{(\alpha)}(\tau) = \frac{\pi(\xi_\tau)}{\Tree_\alpha(\xi_\tau)^{\gamma-1}} m^{(\alpha)}_{\xi_\tau}(\tau),
\end{align*}
as defined above in \eqref{eq:equivnualphas} and the measure expresses the probability of a partition multiplied by $m^{(\alpha)}_\xi$, which is the probability of seeing a forest, $\tau$, conditioned on a partition with $\tau\in \xi_\tau \subset \mathcal{F}_d(G)$. 
 As discussed at the end of the 1-Tree Cycle Walk section, $m^{(\alpha)}_\xi$ so constructed is an invariant measure for the 1-Tree Cycle Walk. This means that there is no need to Metropolize the 1-Tree Cycle Walk because $Q_{\treeOne}( \tau, \xi_\tau)=1$ for any $\tau \in \xi_\tau$ and $\mu_\alpha Q_{\treeOne}= \mu_\alpha$. Hence in our setting, we will take  $P_\treeOne=Q_\treeOne$ and $\nu_\gamma^{(\alpha)}$ is a stationary measure of $P_\treeOne$ for any $\gamma \in \R$. In most of our examples, we will take the weights $\alpha$ and $\beta$, used in defining respectively $Q_\treeOne$ and $Q_\treeTwo$, to be constant.  This will imply that the measure obtained by restricting $\nu$ to any set of the form $\{ \tau' : \xi_{\tau'}= \xi_\tau\}$ for some $\tau \in \mathcal{F}_d$ will be uniform. In the one case where we do weight the graph, we will take the weights to be equivalent for the 1- and 2-tree Cycle Walks, meaning we set $\alpha = \beta$. In implementing the 2-tree Cycle Walk, our rejection probabilities take on the form given in \eqref{eq:accept2treegeneral}.

\section{Algorithmic Considerations}
We store the forest using a linked-cut tree data structure which enables us to add and remove cycles quickly as described in \cite{russo2018linking}. We utilize the splay tree structures with the linked-cut tree structures which provide amortized run times of $O(\log(n))$ for `accessing' (also called `exposing') a node, adding or removing an edge from the tree (i.e., linking and cutting), changing the root of a tree, and finding the root of a node's tree. 

In designing our algorithm, we must be able to perform a depth-first search in order to determine the populations or weights on each side of any edge (as described in \cite{autry2023metropolized}). Link-cut trees traditionally store parent, but not child information, so we augment the structure by adding child information to each node. Modifying child information will be at worst $O(n)$, but because each node will only be touched $O(1)$ times in the listed operations above (depth first search, link/cut, find root, etc.) this will add an amortized constant number of operations.

The 1-Tree Cycle Walk is performed as in \cite{russo2018linking}, with the exception that we ensure the edge we select has its nodes in the same tree (by finding the roots of both nodes). The outline of the algorithm is that once we have sampled an appropriate edge, we re-root the tree with one of the nodes, find the path from the other node to the new root, randomly remove one of the edges along the path, and link the selected edge. 

For the 2-Tree Cycle Walk, we store a list of all cross-district edges, hashed by district pair. We select a random adjacent district pair and then sample two edges to form a loop.  This forms a cycle; each node in the cycle has a certain mass associated with itself and leaves which we compute via a depth-first search in which we track the first and last time we enter the node in order to determine the accumulated mass associated with it along the cycle (as described in \cite{autry2023metropolized}).  We then determine all possible pairs of edges we could remove to arrive at a new balanced forest and select a pair at random. 

\section{Numerical Experiments}\label{sec:NumExp}
We now present a number of numerical explorations of the Metropolized Cycle Walk. We begin by validating the mathematical algorithm presented as well as our implementation on two test cases. The first validation example is a simple, small graph whose symmetries allow it to be solved analytically.  The second validation is a large, realistic problem of public policy interest and for which we have already sampled the chosen measure using a different algorithm, namely Metropolized Forest Recombination. After the validation experiment, we then turn to a number of numerical experiments to explore the convergence of the algorithm to the desired stationary measure. We explore the rate of convergence to a number of different target measures, with different choices of the parameters used in defining the algorithm. Lastly, we dive deeper into properties of the Metropolized Cycle Walk and try to use them to understand some of the behavior we saw in the previous numerical experiments. In particular, we try to understand why Metropolized Cycle Walk seems to perform better in some respects when compared to  Metropolized Forest Recombination.

\subsection{Validations}
We validate our methods in two ways. First, we validate on a small enumerable example and compare an empirical observable with an exactly computed result. Second, we sample from a measure on a larger graph, the precinct graph on North Carolina. In this case, we test for convergence across multiple chains on a certain observable within an algorithm and then compare the current method with that presented in \cite{autry2021metropolized} for a single level of the hierarchy. In the latter case, we are more limited in the measures we can investigate since (i) there is no known way to enumerate all of the possible districts in larger graphs and (ii) the RECOM variants that target known measures have a limited class of measures they can efficiently sample.

\subsubsection{Validation on a $4\times4$ Grid}
We begin by testing our methods against a small graph with an enumerable state space. To do so we examine 4 districts on a $4\times4$ grid (see Fig.~\ref{sfig:4x4setup}). We assume equal population across all 16 nodes and each district to be comprised of exactly 4 nodes. 

We begin by testing that our sampler recovers the $\nu_0$ and $\nu_1$ measures on the uniform grid. We examine the fraction of the time we see a partition with a certain number of cut edges and compare it to the exact probability calculated from the enumerated result. 

We then modify the boundary lengths between the nodes to test modifying the score function, $J$, to account for the isoperimetric ratios of the districts. The isoperimetric ratio is proportional to the inverse of the Polsby-Popper score, which is a traditional measure of compactness in redistricting. To determine the isoperimetric ratios of the districts, we must be able to calculate the perimeter and area of each district. We must therefore add information to the graph, including perimeter lengths between nodes, total nodal perimeter lengths, and node areas.

We set the total perimeters and shared perimeter lengths between nodes so that the score of any given districting plan will remain invariant under rotation and mirror symmetry as shown in Figure~\ref{fig:4x4sym}. This ensures that under rotation or flips, a given district has the same probability. As shown in Figure~\ref{sfig:dualsym}, we modify the lengths between the boundaries so that the edges on the outer corners of the dual graph have a length of 2,  those on the outside middle have a length of $2.5$, and those in the inner box have a length of 3. The remaining edges have length 1.
The nodes on the corners are set to have an additional perimeter of 2, and those in the middle of the exterior 1. The areas of each node are set to be 1. We remark that `wiggles' in the straight gray lines shown on the gray regions in Fig.~\ref{sfig:4x4setup} would allow us to physically achieve these values; therefore, the example, although contrived, has not been abstracted beyond our original intent.

The score function is defined as a weighted sum of the isoperimetric ratios. Greater isoperimetric ratios imply greater district boundaries for a given area; hence, the  more probable districts have smaller perimeters. The score, $J$, is then defined as 
\begin{align}
J(\xi) = J_{\Compact} = w_{\Compact} \sum_{i=1}^4 \frac{(\text{perimeter}(D_i(\xi)))^2}{\text{area}(D_i(\xi))},
\label{eq:compactscore}
\end{align}
where $\text{perimeter}\big(D_i(\xi)\big)$ and $\text{area}\big(D_i(\xi)\big)$ is the perimeter and area of district $i$, respectively, and $w_{\Compact}$ is a weight. In the test, we set $w_{\Compact}=0.02$.

Finally, we repeat the above test by changing the edge weights of the graph, rather than adding perimeter and area information to alter the relative probability mass associated with each forest. In this case, the probability of seeing a particular district is proportional to the sum of the product of edge-weights in each partition as described above.

In all three cases, we examine the fraction of the time we see a partition with a certain number of cut edges and compare it to the exact probability calculated from the re-weighted enumerated result. All tests are present in the test suite of our code repository.\footnote{The tests and Cycle Walk code can be found at \url{https://github.com/jonmjonm/CycleWalk.jl.git}. However, the CycleWalk.jl package is also a registered Julia package; and, hence, can be installed directly into Julia. See Section~\ref{sec:codeBase} in the Appendix. Example of  running Cycle Walk can be found at \url{https://quantifyinggerrymandering.pages.oit.duke.edu/codedoc/}}

\begin{figure}[!htbp]
\centering
\begin{subfigure}[T]{0.49\textwidth}
\centering
\begin{tikzpicture}[tock/.style={decorate, decoration={snake, amplitude=1pt, segment length=3pt}, thick}]

  \tikzstyle{tick} = [{line width=0.3pt}, -]

  \foreach \x in {-0.5,0.5,1.5,2.5,3.5} {
    \foreach \y in {-0.5,0.5,1.5,2.5} {
      \draw[customgray] (\x,\y) -- (\x,\y+1); 
    }
  }

  \foreach \y in {-0.5,0.5,1.5,2.5,3.5} {
    \foreach \x in {-0.5,0.5,1.5,2.5} {
      \draw[customgray] (\x,\y) -- (\x+1,\y); 
    }
  }

  \foreach \x in {0,1,2,3} {
    \foreach \y in {0,1,2} {
      \draw (\x,\y) -- (\x,\y+1); 
    }
  }

  \foreach \y in {0,1,2,3} {
    \foreach \x in {0,1,2} {
      \draw (\x,\y) -- (\x+1,\y); 
    }
  }

  \foreach \x in {0,1,2,3} {
    \foreach \y in {0,1,2,3} {
      \filldraw[black] (\x,\y) circle (2pt);
    }
  }
\end{tikzpicture}
\caption{Precinct graph in light gray and Adjacency/graph dual graph in black with nodes for each precinct.}
\label{sfig:4x4setup}
\end{subfigure}\hspace{0.5cm}
\begin{subfigure}[T]{0.45\textwidth}
\centering
\begin{tikzpicture}[tock/.style={decorate, decoration={snake, amplitude=1pt, segment length=3pt}, thick}]
  \tikzstyle{tick} = [{line width=0.3pt}, -]

  \foreach \x in {-0.5,0.5,1.5,2.5,3.5} {
    \foreach \y in {-0.5,0.5,1.5,2.5} {
      \draw[customgray, opacity=0] (\x,\y) -- (\x,\y+1); 
    }
  }

  \foreach \y in {-0.5,0.5,1.5,2.5,3.5} {
    \foreach \x in {-0.5,0.5,1.5,2.5} {
      \draw[customgray, opacity=0] (\x,\y) -- (\x+1,\y); 
    }
  }

  \foreach \x in {0,3} {
    \foreach \y in {0,2} {
      \draw (\x,\y) -- (\x,\y+1); 
      \draw[tick] (\x-0.1,\y+0.5) -- (\x+0.1,\y+0.5);
    }
  }

  \foreach \x in {0,2} {
    \foreach \y in {0,3} {
      \draw (\x,\y) -- (\x+1,\y); 
      \draw[tick] (\x+0.5,\y-0.1) -- (\x+0.5,\y+0.1);
    }
  }

  \foreach \x in {1,2} {
    \foreach \y in {0,2} {
      \draw (\x,\y) -- (\x,\y+1); 
      \draw[tick] (\x-0.1,\y+0.475) -- (\x+0.1,\y+0.475);
      \draw[tick] (\x-0.1,\y+0.525) -- (\x+0.1,\y+0.525);
    }
  }

  \foreach \x in {0,2} {
    \foreach \y in {1,2} {
      \draw (\x,\y) -- (\x+1,\y); 
      \draw[tick] (\x+0.475,\y-0.1) -- (\x+0.475,\y+0.1);
      \draw[tick] (\x+0.525,\y-0.1) -- (\x+0.525,\y+0.1);
    }
  }

  \foreach \x in {1} {
    \foreach \y in {0,3} {
      \draw (\x,\y) -- (\x+1,\y); 
      \draw[tick] (\x+0.45,\y-0.1) -- (\x+0.45,\y+0.1);
      \draw[tick] (\x+0.55,\y-0.1) -- (\x+0.55,\y+0.1);
      \draw[tick] (\x+0.5,\y-0.1) -- (\x+0.5,\y+0.1);
    }
  }

  \foreach \x in {0,3} {
    \foreach \y in {1} {
      \draw (\x,\y) -- (\x,\y+1); 
      \draw[tick] (\x-0.1,\y+0.45) -- (\x+0.1,\y+0.45);
      \draw[tick] (\x-0.1,\y+0.55) -- (\x+0.1,\y+0.55);
      \draw[tick] (\x-0.1,\y+0.5) -- (\x+0.1,\y+0.5);
    }
  }

  \foreach \x in {1} {
    \foreach \y in {1,2} {
      \draw (\x,\y) -- (\x+1,\y); 
      \draw[tock] (\x+0.5,\y-0.1) -- (\x+0.5,\y+0.1);
    }
  }

  \foreach \x in {1,2} {
    \foreach \y in {1} {
      \draw (\x,\y) -- (\x,\y+1); 
      \draw[tock] (\x-0.1,\y+0.5) -- (\x+0.1,\y+0.5);
    }
  }

  \foreach \x in {0,1,2,3} {
    \foreach \y in {0,1,2,3} {
      \filldraw[black] (\x,\y) circle (2pt);
    }
  }
\end{tikzpicture}
\caption{Adjacency Graph/Dual graph edge lengths marked: single tick  $|$ is 2,  double tick  $||$ is 1, triple tick $|||$ is 2.5, and squiggly tick $\sim$ is 3.}
\label{sfig:dualsym}
\end{subfigure}
\caption{}
\label{fig:4x4sym}
\end{figure}

\subsubsection{Validation on a Larger Graph}
We now validate our method on a larger graph by comparing the measure obtained by sampling $\nu_\gamma$ with Metropolized Forest Recombination to those obtained by sampling using the Metropolized Cycle Walk algorithm described in this note. We sample  $\nu_0$ and  $\nu_{0.2}$ on the precincts graph of North Carolina and a 2\% population deviation on the 2020 census data to create redistricting plans with 14 districts (equal to the number of the Congressional districts in 2020). 
We chose $\gamma=0$ because it is what RECOM appears to sample efficiently. We chose $\gamma=0.2$ because RECOM appeared to still sample well in this regime, allowing us to compare Cycle Walk at a different target measure.
We launch 4 independent chains for each method. We choose the  Cycle Walk parameters so, on average, there are  99  \treeOne Cycle Walk steps for every \treeTwo Cycle Walk step.

\begin{figure}[!htbp]
\centering
\includegraphics[width=0.48\textwidth]{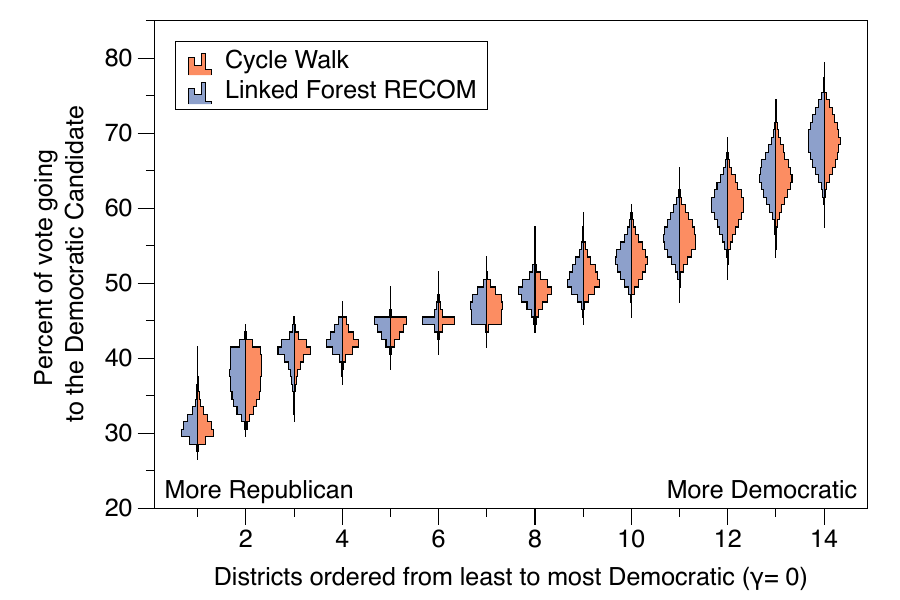}
\includegraphics[width=0.48\textwidth]{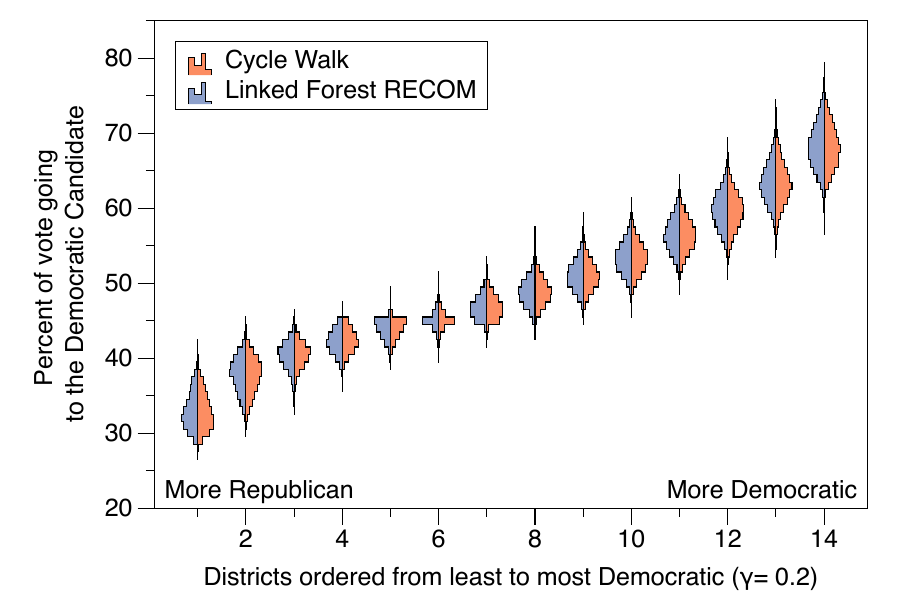}
\newline
\includegraphics[width=0.48\textwidth]{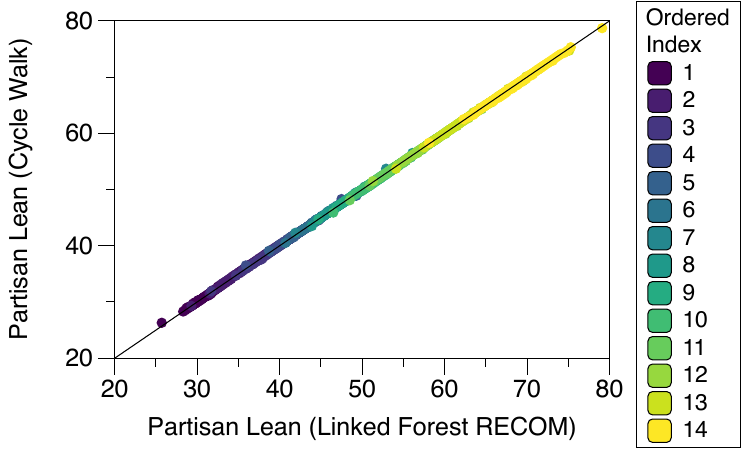}
\includegraphics[width=0.48\textwidth]{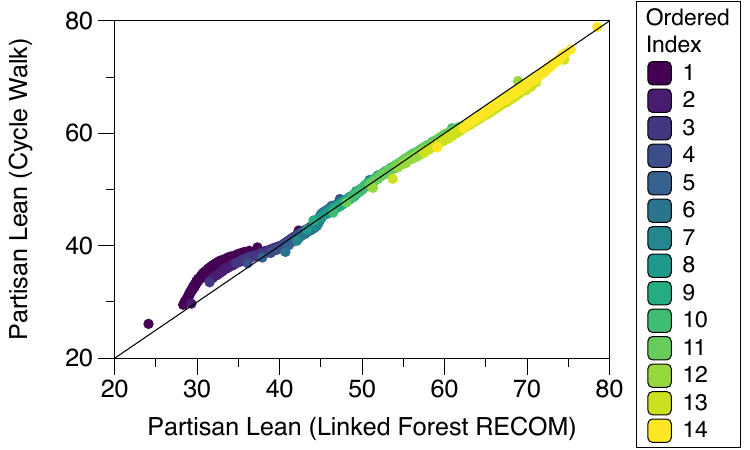}
\caption{Comparison of the rank-ordered marginals from the most Republican to the most Democratic district over samples of $\nu_0$ (left) and $\nu_{0.2}$ (right) applied to North Carolina using the votes for the 2020 presidential election. On top, the orange gives the marginals using samples generated by the Cycle Walk algorithm and the blue using samples generated by the Metropolized Forest Recombination algorithm. On bottom, we display the QQ plots for each marginal distribution by color.}
\label{fig:validateNC}
\end{figure}

In both cases, we use the 2020 Presidential vote counts to calculate the Democratic vote fraction in the $k$th most Republican district for each redistricting plan generated through a run of the algorithm. By letting $k$ vary over $\{1,2,\dots,14\}$ we obtain 14 different marginal histograms. These 14 histograms are compared in \Cref{fig:validateNC} for the Cycle Walk (in orange) and  Forest Recombination (in blue) algorithm, both Metropolized to sample from either  $\nu_0$ or  $\nu_{0.2}$. The results show that the two algorithms appear to be generating the same marginal distributions. We also display the QQ plots to compare each marginal: The marginal distributions from both figures line up well, with the exception of some  variation off of the diagonal for the two most Republican districts at $\gamma = 0.2$.

To validate the convergence properties within the methods, we examine the Gelman-Rubin convergence statistic for each $k$th most Republican district across the 4 runs for each method. For the Cycle Walk, we find an average Gelman-Rubin statistic of 1.0000072 and 1.000029 and a maximum of 1.00018 and 1.00033 across the 14 observables for $\gamma=0$ and $0.2$, respectively, after $10^9$ total proposals (and roughly $10^7$ Cycle Walk proposals).  For the Metropolized Forest Recombination, we find an average Gelman-Rubin statistic of 1.00026 and 1.0033, and a maximum of 1.0010 and 1.0073 across the 14 observables for $\gamma=0$ and $0.2$, respectively, after 950,000 total proposals.

\subsection{Studies of Convergence Rates and Mixing Times}\label{sec:ConvergenceRates}

We will now give a number of different convergence studies. Generally, we will again compare the ranked-ordered marginals of the $k$th most Republican district, where $k$ varies from one to the total number of districts in the plan. We will consider congressional district plans from North Carolina and Connecticut, where the number of districts in each plan is 14 and 5, respectively. We will also consider a few synthetic graphs where the number of districts will vary. We will explore the convergence rate by launching a number of runs from different random initial district plans. We then compare the total variation distance between the empirical measure of the above marginals obtained from different initial conditions. We report the maximum distance when all pairs of initial data are compared. If all of the trajectories starting from different initial conditions produce similar empirical measures, then it is a strong indication that the chains have converged to a unique statistical state. Of course, it is possible that the state space has only reached a local equilibrium with a very unlikely transition needed to reach an unexplored part of the configuration space. However, such statistics are widely accepted in applied statistical settings requiring Markov chain sampling. While we did not measure distance between the full empirical measure but rather only a collection of marginals, this choice is very defensible: The chosen marginals can be used to answer many of the most important policy questions in the redistricting context.

\subsubsection{Comparison of Convergence Rates while Interpolating between Uniform Distribution on Partitions and on Forests for North Carolina}\label{sec:ratesGammaNC}

We explore the convergence rate for the sampling of $\nu_\gamma$ as $\gamma$ ranges from zero to one. Recall that when $\gamma=0$ and $J\equiv0$ the measure is uniform on spanning  forests in $\mathcal{F}_d$ and when $\gamma=1$ and $J\equiv0$ the measure is uniform on partitions with $d$ elements. The log-log plots show the convergence of the average deviation of the marginal empirical measures over the range of $\gamma$. As expected, the convergence rate degrades as $\gamma$ approaches 1, both for the Metropolized Cycle Walk and for the Metropolized Forest Recombination algorithms. However, the  Metropolized Cycle Walk shows a reasonable level of convergence for much higher $\gamma$ when compared to  Metropolized Forest Recombination. While this result merits more investigation to fully understand, it does underline an expected strength of the  Metropolized Cycle Walk over the Metropolized Forest Recombination. As we will see in \Cref{sec:structureCycleWalk}, the Cycle Walk proposal does not replace the entire spanning tree with a new tree; it is more likely to make the small and medium-sized moves that are in keeping with a wider range of target measures.  Furthermore, unlike the single-node flip of a standard Swendsen–Wang proposal \cite{SwendsenWang,fifield2020automated}, the chain has approximate population balance baked into the proposal.

\begin{figure}[!htbp]
\centering
\includegraphics[width=\textwidth]{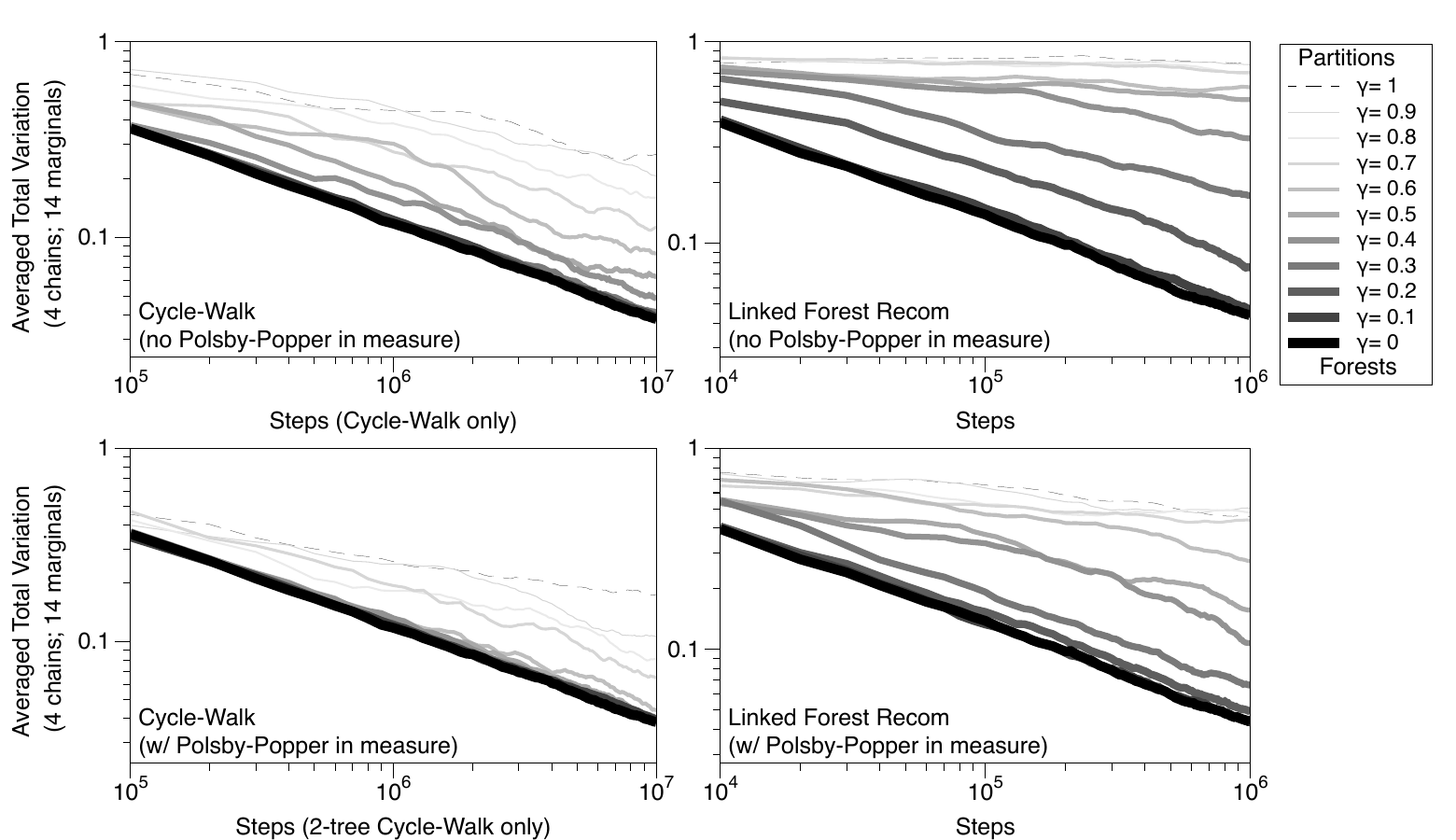}
\caption{We examine the average error between rank-ordered marginals of the least to most Democratic districts under the Presidential 2020 votes on 14 North Carolina congressional districts. The error is averaged across 4 independent runs as a function of the number of proposals. We display the Cycle Walk (left) and Linked Forest Recombination (right) across different measures $\nu_\gamma$ for a range of $\gamma$ values. On the top, we show the convergence rates with a score that does not account for the Polsby-Popper score; on the bottom, we tune a Polsby-Popper score with $\gamma$ target (roughly) plans with similar compactness scores (see \Cref{fig:isoperimetric}).}
\label{fig:convergenceWGammaStudy}
\end{figure}

We begin by examining the relative convergence properties as a function of the measure. We start with the simplest measure, using $\nu_\gamma$ with $\gamma \in [0,1]$. We select a number of discrete choices for $\gamma$, including $\gamma=0$, $\gamma=1$, and several intermediate values. For each value of $\gamma$, we run 4 independent Markov chains using the Cycle Walk algorithm. We parametrize our runs by taking on average 99 \treeOne Cycle Walk steps (non-Metropolized) for every one \treeTwo Cycle Walk steps, and running for a total of $10^9$ proposals (or $10^7$ expected 2-tree Cycle Walk steps). We run our chains on 14 districts of the 2021 North Carolina precincts with a 2\% population deviation under the 2020 census data and output every 1000 steps 2-tree Cycle Walk steps. All of the independent chains on North Carolina were run on an AMD Ryzen 9 5950X 16-Core Processor and typically took between 13 and 40 hours to run on $10^7$ 2-Tree Cycle steps.\footnote{There was a high variance to this; even with the same parameters, some of the runs could take twice as long as the others. A small fraction of the runs took just over 3 days to complete, even when other runs with the same parameters took less time. This may have been due to issues with particular nodes on our local cluster.} 

We compare the Cycle Walk method with independent chains that ran $10^6$ steps of linked-Forest Recombination (see \cite{autry2021metropolized}) and repeat our experiments using this algorithm. 
When selecting the number of steps to run the chains, we initially reasoned that an accepted Cycle Walk step would, in general, lead to less change than an accepted step in RECOM. We thus ran Cycle Walk for 10 times more steps.

We expect there to be fast convergence properties for low $\gamma$ since the uniform spanning forest measure does not alter the Metropolis-Hastings acceptance ratio. We therefore use $\gamma=0$ as a benchmark. Furthermore, we expect to see poor convergence results at $\gamma=1$ since (i) we expect typical partitions to be fractal space-filling-curve-like objects (see \cite{najt2019complexitygeometrysamplingconnected, najt2021empirical}) and (ii) this, in some sense, has been the sought-after but difficult base-measure to access in this field.  We confirm these assumptions and display our results in the top row \Cref{fig:convergenceWGammaStudy} in which we see for both the Cycle Walk (left) and linked-Forest Recombination (right) that the convergence is fastest for small $\gamma$ and we see no signs of convergence at $\gamma=1$. The point at which convergence properties break down as $\gamma$ grows, however, differs for the two algorithms. For linked-Forest Recombination, the convergence is similar for $\gamma\in[0,0.1]$ but then quickly begins to slow down; by the time $\gamma=0.3$ the average error in the ranked-ordered Marginals is over 4 times that of $\gamma=0$ after the same number of steps in the Markov chain. In contrast, there is no significant difference in the average error for the Cycle Walk algorithm between $\gamma=0$ and $\gamma=0.3$; by the time $\gamma=0.5$, the error has grown by only about 50\%. 

We further validate these observations by estimating the effective independent sample rate. We do this in the typical way by summing the autocorrelation function for each of the rank-ordered marginal distributions on the partisan composition as described above. We report the number of samples we need to take in order to, in effect, get a single independent sample which is defined as $1 + 2 \sum_{i=1}^\infty \rho_i$, where $\rho_i$ is the autocorrelation function after $i$ steps in the chain, and where we truncate the sum to when we first see $\rho_i$ go below zero, i.e. $i_{-} = \min{i | i>0, \rho_i \leq 0}$. We only examine a single chain per value of $\gamma$; for $\gamma$ values at which the chains have not converged, these are lower bounds on the number of samples needed in the Markov chains to take an effective independent sample. Finally, we remark that we take a 2-Cycle Walk step with only 1\% probability (and otherwise take a 1-Cycle Walk step), however, we report the effective sample rate based on the expected number of 2-Cycle Walk steps.

\begin{figure}
  \centering 
\includegraphics[width=0.5\textwidth]{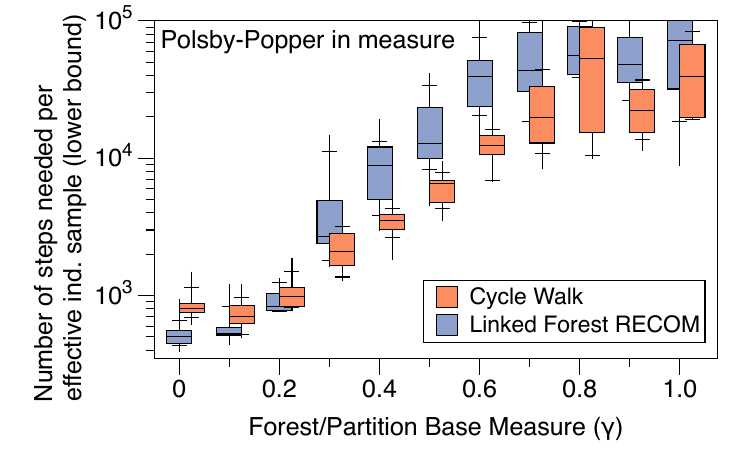}
\caption{We display the range of effective sample rates across the 14 ranked marginal distributions of the partisan makeup (most Republican to most Democratic) under the 2020 Presidential voting data. For each value of $\gamma$ considered, we display a box plot to show the variation in effective sample rates. We display our results when explicitly controlling for the isoperimetric ratio in the measure (which directly relates to the Polsby Popper compactness score). The boxes represent the quartiles; the upper and lower tick marks represent the 10th percentiles (or demarcate the slowest/fastest 2 of the 14 observables).}
\label{fig:ess}
\end{figure}

We examine the same range of $\gamma$'s; for each value, we look at the 14 effective sample rates and display them as a box plot for the Cycle Walk and linked-Forest Recombination algorithms (see Figure~\ref{fig:ess}). We find that at $\gamma=0$, linked-Forest Recombination has a faster effective sample rate, needing roughly half the samples as Cycle Walk. However, as $\gamma$ increases, this quickly changes. At $\gamma=0.2$ the rates are comparable and at $\gamma=0.3$ and above, Cycle Walk requires fewer steps to achieve the same number of effectively independent samples. In short, the Cycle Walk algorithm is a much more efficient sampler for larger values of $\gamma$.

Comparing the per-step computational expense of Cycle Walk to linked-Forest Recombination depends on the ratio of 1- to 2-Cycle Walk steps, and the expense of uniformly sampling a random spanning tree in Recombination. In general, we expect making several 1-Tree Cycle Walk steps to be less expensive than using Wilson's algorithm (and Wilson's happens at every step of Recombination). Although we forgo a formal analysis, we expect that the overall computational cost of gathering effective samples in Cycle Walk will be less than linked-Forest Recombination for all values of $\gamma$ studied above.

While this is an interesting finding, the target measure is not of particular interest in the gerrymandering context as $\gamma$ grows. This is due to the fact that the measure increasingly becomes concentrated on serpentine districts that would approach space-filling as the lattice is refined. To keep the measure concentrated on plans that are relatively compact, we will introduce a score term $J_{\Compact}$ which is the sum of the isoperimetric constants for each of the districts in the plan. We will then consider a sequence of measures $\nu_\gamma(\tau)$ as defined in \eqref{eq:nudefinition}
with the exponent given by  $- \gamma J_{\Tree}(\xi_\tau) - c_\gamma J_{\Compact}(\xi_\tau)$. We will choose $c_\gamma$ so that the sum of the isoperimetric ratios varies over a similar range for each measure as it did for $\nu_0$. The validation of this tuning is given in \Cref{fig:isoperimetric}. We observe a (roughly) linear scaling in $c_\gamma$ and report the constants in Appendix~\ref{apdx:tuning_compactness_weights}
\begin{figure}[!htbp]
\centering 
\includegraphics[width=0.48\textwidth]{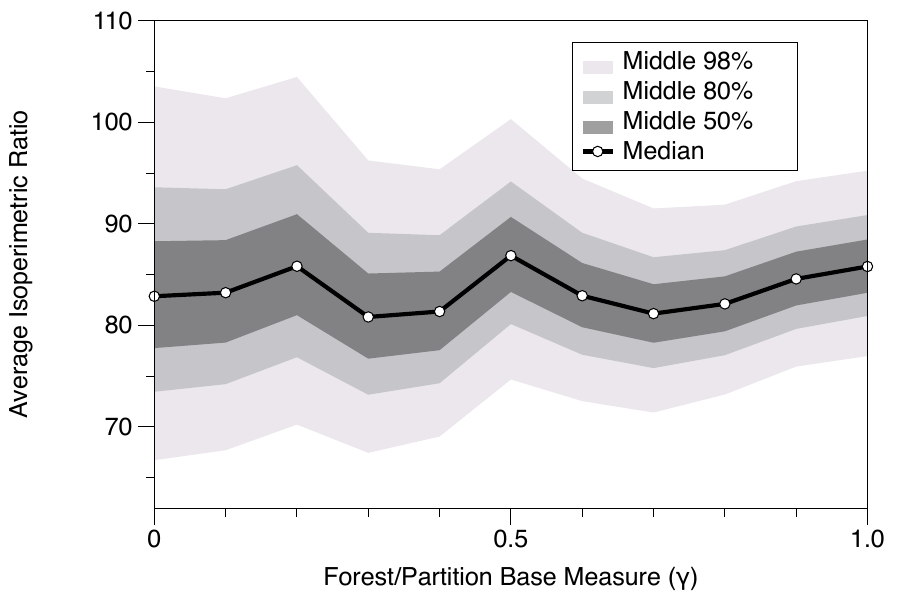}
\includegraphics[width=0.48\textwidth]{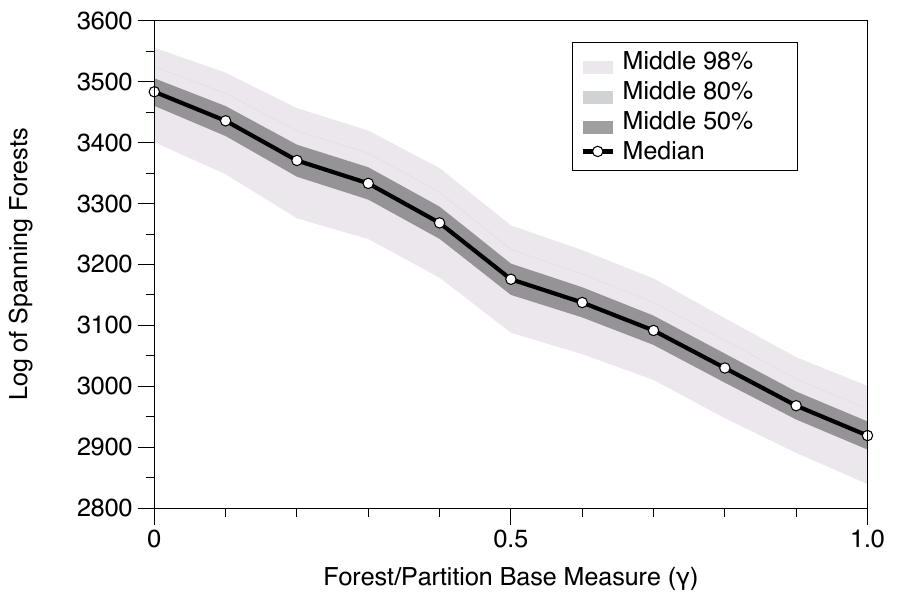}
\caption{Quantiles of compactness (left) and forests (right) as we tune the weight of the sum of isoperimetric scores in the measure. We have tuned the weights so that the average isoperimetric is roughly constant (as shown in the left plot). A further refinement of the weights could produce a straighter line on the left figure, but this is sufficient for our current purposes. The figure on the right shows how the samples focus on different ranges of spanning forests.}
\label{fig:isoperimetric}
\end{figure}

In the bottom row of \Cref{fig:convergenceWGammaStudy}, we give the analogous plot as before, but now with the compactness constrained through the isoperimetric score. Notice that the Metropolized Cycle Walk performs even better when compared to Metropolized Forest Recombination as $\gamma$ grows. The Cycle Walk convergence properties seem to be essentially equivalent up through $\gamma=0.6$, whereas the linked-Forest Recombination algorithm begins to slow even as early as $\gamma=0.2$.

\begin{remark}
The above results show a promising structure of the error as we diverge from the spanning forest measure. We have also measured the wall-clock time using the BenchmarkTools in Julia. Testing at $\gamma = 0.1$ with an isoperimetric weight activated on North Carolina, we find that Metropolized Forest Recombination takes about 0.011 seconds per step. In contrast, taking a single \treeTwo Cycle Walk step for every 99 \treeOne Cycle Walk steps, the algorithm takes 0.0036 seconds per \treeTwo Cycle Walk step; using a ratio of a single \treeTwo Cycle Walk step for every 9 \treeOne Cycle Walk steps (the optimal found below), the code accelerates to 0.0022 seconds per \treeTwo Cycle Walk step.

In short, using the optimal ratio (discovered in the next section), the Cycle Walk algorithm runs roughly 5 times faster per state-changing step than Forest Metropolized Recombination. The overall speed of the algorithm (in achieving some desired error) will depend on the measure we sample from.
\end{remark}

\subsubsection{Comparison of Convergence Rates while Varying Ratio of 1-Tree Cycle Walks to 2-Tree Cycle Walks}
We now explore the effect of varying the fraction of 1-Tree Cycle Walks to 2-Tree Cycle Walks.  It is not clear \emph{a priori} what the effect will be or if it will be the same across different choices of target measure or graphs. As already mentioned, the 1-Tree Cycle Walk will at times be referred to as the ``Internal Cycle Walk''. This is because it is internal to the partitions implied by the trees of the spanning forest. The partition is constant if only one of the trees changes, hence the 1-Tree Cycle Walk does not change the implied partition.  Hence, if the score function that determines the target measure, through \eqref{eq:pi_def}, is constant on the set of spanning forests that determine the same partition then the 1-Tree Cycle Walk will not change the score. In other words, alone the 1-Tree Cycle Walk evolves on a set of spanning forests that have identical scores and hence equal probability.

Thus, the 1-Tree Cycle Walk alone will sample a connected component of a level set on the score function. In contrast, the 2-tree Cycle Walk does change the partition; and hence, Metropolization will produce a walk that is influenced by the score function and, ideally, samples from the measure the score function defines via \eqref{eq:pi_def}. 
It might be natural to think that the 1-Tree Cycle Walk does little to explore the state space and speed the MHMC towards the desired stationary measure. On the other hand, fluctuations in the underlying tree may provide novel pathways for the 2-Tree Cycle Walk to evolve. We will see that the latter perspective is correct; however, there are diminishing returns in adding fluctuations in the trees.

\begin{figure}[!htbp]
\includegraphics[width=0.48\textwidth]{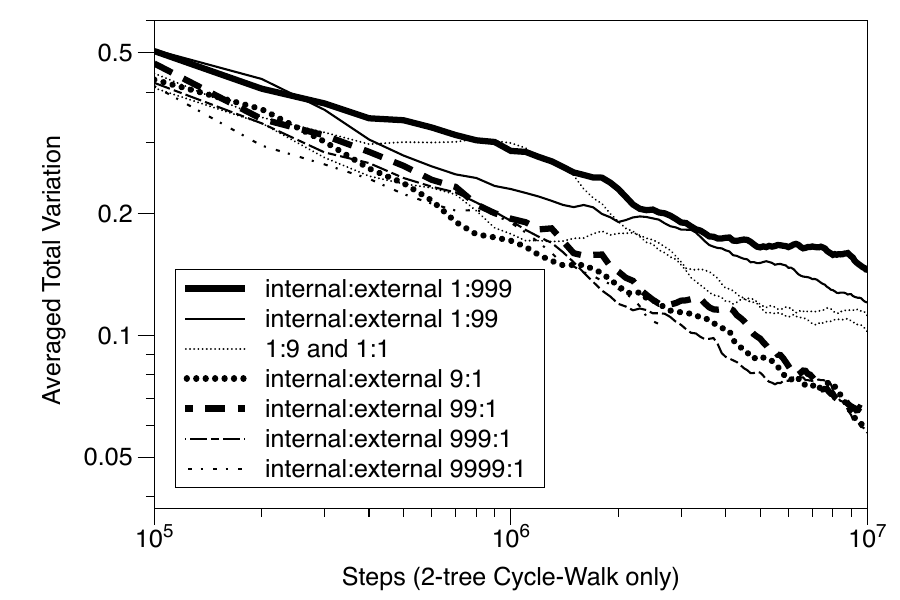}
\includegraphics[width=0.48\textwidth]{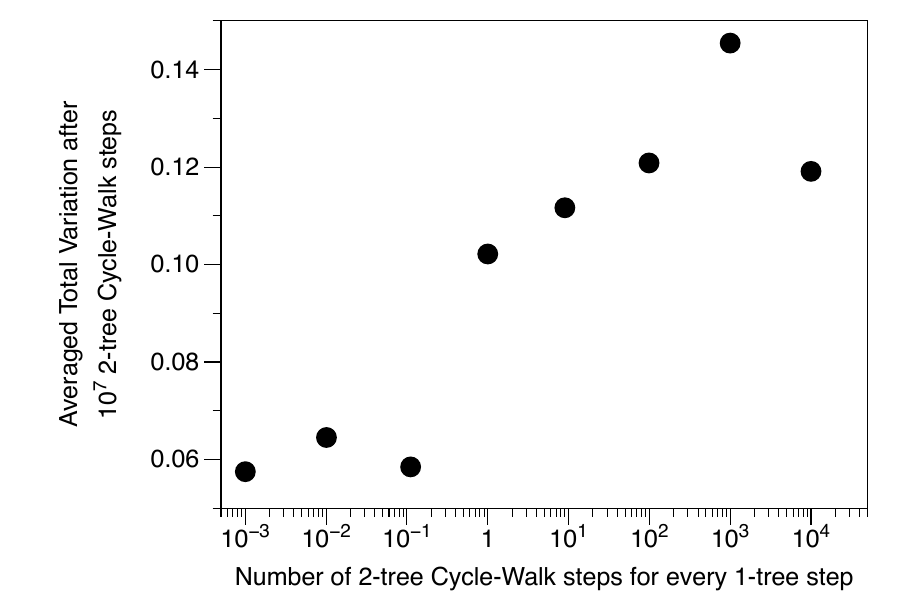}
\caption{The average error in the 14 Ranked Ordered Marginals Congressional district plans using the vote from the 2020  Presidential race. The error is averaged across 4 independent chains. The plots show a range of ratios of 1- vs 2-Tree Cycle Walks. The plot is with $\gamma=0.7$ and all runs include an isoperimetric compactness score with the same weights as in \Cref{fig:isoperimetric}. We show the error as a function of the steps (left) and the error after $10^7$ Cycle Walk steps (right).}
\label{fig:varyInternalRatio}
\end{figure}

In \Cref{fig:varyInternalRatio}, we consider the convergence of the marginal distributions of the rank-ordered marginals of the sampling of a measure on congressional districting plans of North Carolina with an isoperimetric compactness score already considered in  \Cref{fig:isoperimetric}. We consider this measure with $\gamma$ equal to $0.7$ but with a range of different relative frequencies of the 1-Tree Cycle Walk and the 2-Tree Cycle Walk. We recall that on each step of the chain, we randomly chose between the two proposals with the stated relative frequency. This study uses the measure that accounts for the isoperimetric ratio to keep districts more compact. We select $\gamma=0.7$ because this is the lowest value of which we see a divergence in the convergence properties; varying the ratio of step-types allows us to examine if it is possible to improve or degrade convergence. All of the walks are run so the same number of 2-tree Cycle Walk proposals are made. 

Adding 1-Tree Cycle Walks accelerates the convergence to equilibrium. It appears that nine 1-Tree Cycle Walks for each 2-Tree Cycle Walk is a good choice because we do not see improvement by adding more 1-tree Cycle Walk steps, and these types of steps do not make any changes to the associated partition. There are a number of possible explanations for this effect. More investigations are needed to understand, but two possible mechanisms that likely contribute to this improvement of the 9:1 ratio come to mind. First, it is possible that some tree configurations are not advantageous for producing proposals that exchange large to medium numbers of vertices, which are then accepted in the Metropolization step. The 1-Tree Cycle Walk ensures that the tree structure inside each district changes substantially over time.  Secondly, as already mentioned, one strength of the RECOM methods is that each proposal for how to split the two districts being merged is independent of the previous configuration of the districts, conditioned on the other changed districts. This means that each step has a high degree of independence. This is less true of the 2-Tree Cycle Walk on its
own. However, since the 1-Tree Cycle Walk in isolation has the uniform measure on spanning trees as its stationary measure, it can be used to decorrelate internal tree states between successive 2-Tree Cycle Walk proposals. 

\subsubsection{Comparison of Convergence Rates with Multiscale Single-Node Flip Method}
In \cite{chuang2024multiscaleparalleltemperingfast}, there was a hierarchical multiscale method that used single-node flips on a cascade of coarsened graphs. The method was able to achieve efficient mixing on Connecticut's 5 congressional districts over (roughly) 700 precincts on a measure that was weighted by a partition's isoperimetric score and independent of its tree count. 

\begin{figure}
\centering
\includegraphics[width=0.48\textwidth]{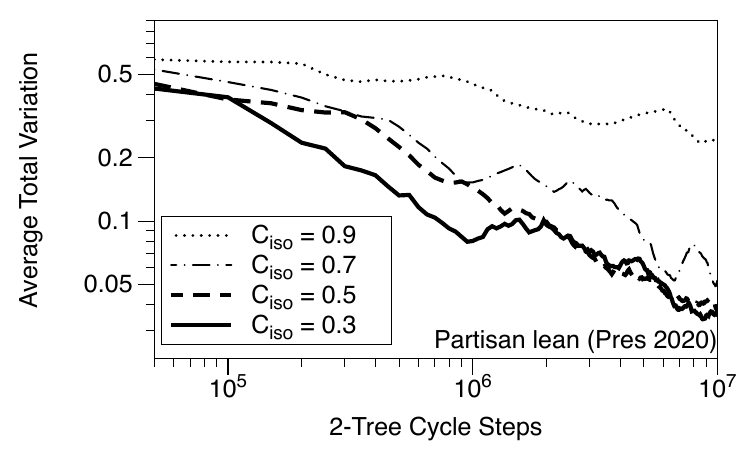}
\includegraphics[width=0.48\textwidth]{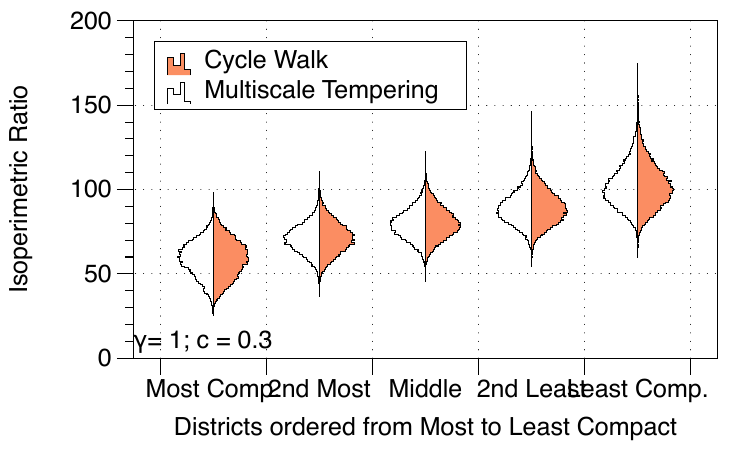}
\includegraphics[width=0.48\textwidth]{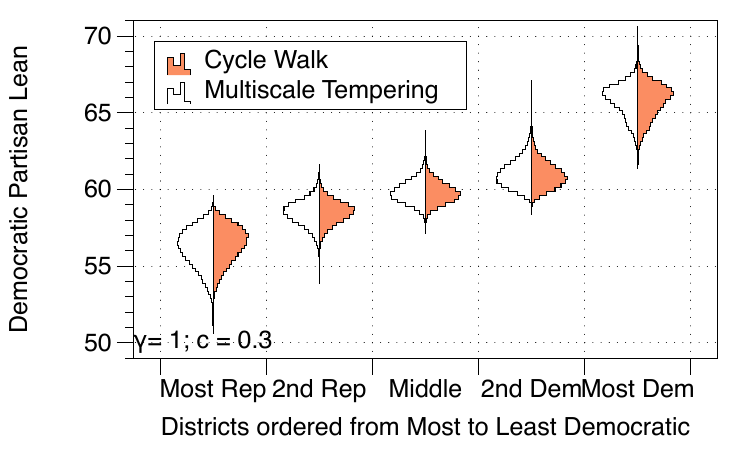}
\caption{We examine the convergence properties of the Cycle Walk on 5 districts within Connecticut with $\gamma=1$ and the weight on the sum of the isoperimetric ratio (as part of the measure) at 0.3, 0.5, 0.7, 0.9. We compare the results from \cite{chuang2024multiscaleparalleltemperingfast} with an isoperimetric weight of $c = 0.3$ against the ordered marginals of isoperimetric ratios and partisan lean according to the 2020 presidential votes.}
\label{fig:CTconvergence}
\end{figure}

We investigate if the Cycle Walk method is similarly able to sample from this space. We again take the exponent in the measure to be ${-J_{\Tree}(\xi_\tau) - c J_{\Compact}}{(\xi_\tau)}$ (with $\gamma = 1$) and set $c=0.3, 0.5, 0.7, 0.9$, which overlaps the values chosen in \cite{chuang2024multiscaleparalleltemperingfast}. We examine the convergence properties for each value of $c$ by taking a 2\% population deviation on the 2020 Connecticut precinct graph; we take (in expectation) 9 1-tree Cycle Walks for every 2-tree Cycle Walk steps and study the convergence properties in Figure~\ref{fig:CTconvergence}. We see that the convergence rate slows as $c$ grows. With the exception of $c=0.9$, the runs appear to have good mixing properties. We remark that although the mixing properties for the larger weight are not as good as those in \cite{chuang2024multiscaleparalleltemperingfast}, that the above runs took 1 2-tree Cycle Walk step for every 9 1-tree Cycle Walk steps and took less than 10 hours to run on a single core, meaning that (i) the computational expense is far less than the other method and (ii) adding tempering between $c=0.7$ and $c=0.9$ may achieve good mixing with less cost at the larger weight. We compare the ordered marginal distributions at $c=0.3$ for the isoperimetric ratios and the partisan lean under the Presidential 2020 votes, and also find good agreement between the two methods.\footnote{Here, Democratic Partisan lean is defined as the number of Democratic votes in a district divided by the sum of Republican and Democratic votes.}

\subsubsection{Comparison of Convergence Rates on  Regular Lattices}\label{sec:ConvergenceRegularLattices}
To this point, we have primarily examined the Cycle Walk algorithm on irregular, realistic test cases. It is possible that the irregularity of the resulting graphs, and perhaps even the particular quirks of the selected graphs, have informed our conclusions. Furthermore, we have not yet built a sense of how various parameters, such as graph size or the number of districts, impact mixing times. 

We therefore examine two regular lattices: square and triangular. We remark that using a triangular lattice for node connectivity may be more realistic. The triangular lattice implies a hexagonal precinct adjacency graph (i.e., the dual graph) in which each precinct has 6 neighbors. In comparison, both the North Carolina and Connecticut precinct graphs have an average neighbor count of 5.7 over all internal nodes (i.e., nodes that do not touch the perimeter of the state). 

\begin{figure}
\centering 
\includegraphics[width=0.48\textwidth]{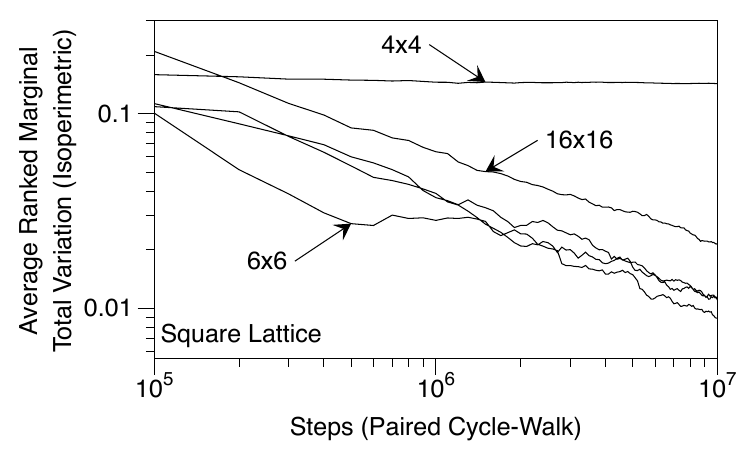}
\includegraphics[width=0.48\textwidth]{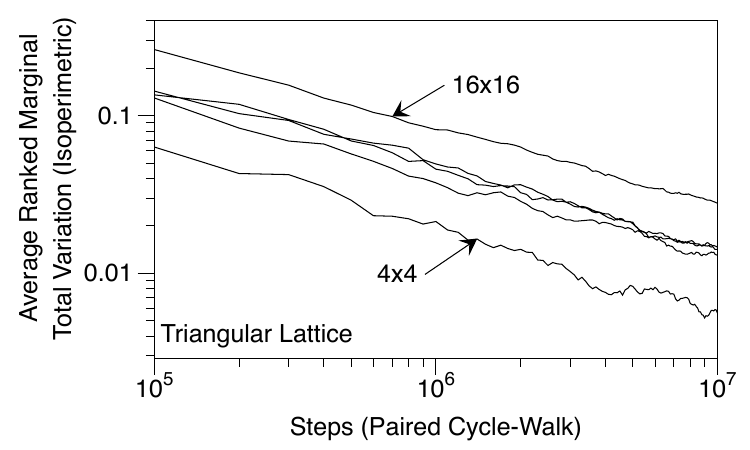}
\caption{We examine the convergence properties of the Cycle Walk on square and triangular lattices of various sizes. We fix $\gamma=0$.}
\label{fig:gridconvergence}
\end{figure}

We begin by sampling $\nu_0$ over a variety of grid sizes. For both square and triangular lattices, we examine grid sizes of $4\times 4$, $6\times 6$, $8\times 8$, $10\times 10$, and $16\times 16$.  We assume that each node has equal (unit) population and set population bounds so that the partitions can vary by a size of one. For example, the $4\times 4$ lattice has an ideal population of $16/5=3.2$, so we ensure that the populations of the districts will be either 3 or 4. We examine the convergence properties as a function of the size of the graph in Figure~\ref{fig:gridconvergence} for 5 districts and see the convergence rate slowing with the size of the graph. Interestingly, the $4\times 4$ square lattice appears to lock in place, and the system is not able to mix, but the larger lattices have the same properties as what we've previously observed. 

\subsection{The Structure and Effectiveness of Cycle Walk Proposals}\label{sec:structureCycleWalk}

We have already seen in \Cref{sec:ratesGammaNC}  and \Cref{sec:ConvergenceRegularLattices}, the Cycle Walk seems able to successfully sample the measure $\nu_\gamma$, with compactness energy tuned to keep the mean compactness relatively constant, for $\gamma$ much closer to one than Metropolized RECOM. It was suggested in the introduction, in \Cref {sec:2Tree}, and above all in \Cref{sec:BeyondRECOM} that this could be explained by the fact that the Metropolized Cycle Walk naturally tunes itself between Metropolized RECOM and Metropolized Single-Node Flip. In the first, every step merges and splits two adjoining districts in a way that has no relation to what was there before. This efficiently injects decorrelation into the Markov chain. In the second, small moves largely based on the current state are made. This allows the state to evolve more smoothly and, when Metropolized, be informed by the structure of the measure more easily.  However, it tends to move more diffusively and has trouble making the big global moves that allow RECOM to mix well. We will now explore the size of the moves made by the Metropolized Cycle Walk when sampling $\nu_\gamma$ under a range of $\gamma$. We will always pick settings and  $\gamma$ for which we have good numerical evidence that the Metropolized Cycle Walk converges. We will see how the distribution of size of accepted proposals changes as $\gamma$ is varied. We will also explore how the acceptance probability varies as the size of the proposed change. Of course, RECOM proposals also have the advantage over Single-Node Flip that the proposals have balanced populations by construction. Cycle Walk shares this property with RECOM.
\begin{figure}[htbp]
  \centering
\begin{subfigure}{0.49\textwidth}
\centering
 \includegraphics[width=0.95\textwidth]{{"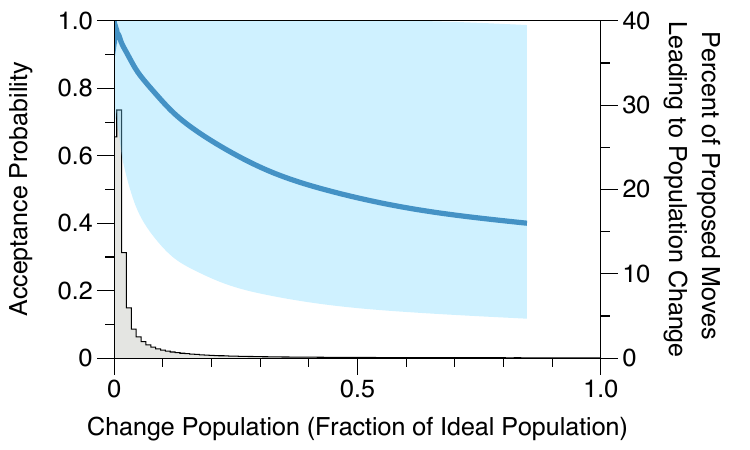"}}   
\caption{$\gamma=0.1$}
\label{fig:acceptanceProbVsDeltaPopFrac_gamma0.1}
\end{subfigure}
\begin{subfigure}{0.49\textwidth}
\centering
 \includegraphics[width=0.95\textwidth]{{"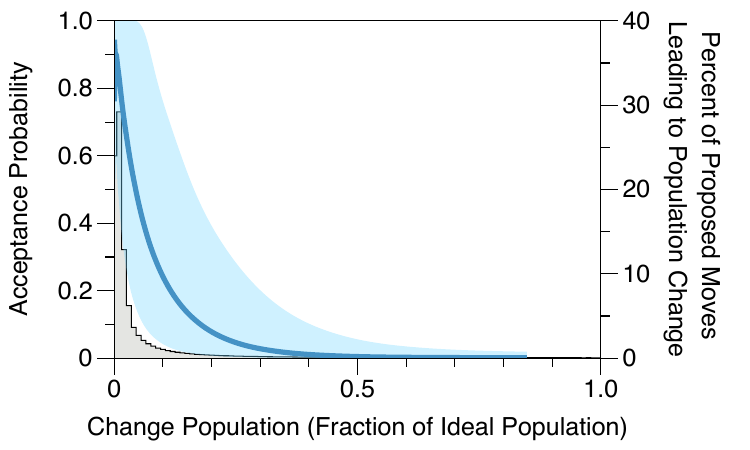"}}
\caption{$\gamma=0.4$}
\label{fig:acceptanceProbVsDeltaPopFrac_gamma0.4}
\end{subfigure}
\\
\begin{subfigure}{0.49\textwidth}
\centering
 \includegraphics[width=0.95\textwidth]{{"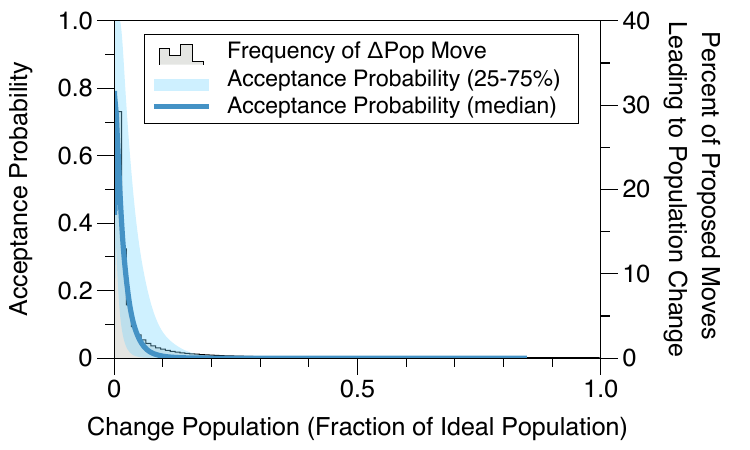"}}
\caption{$\gamma=0.7$}
\label{fig:acceptanceProbVsDeltaPopFrac_gamma0.7}
\end{subfigure}
\begin{subfigure}{0.49\textwidth}
\centering
 \includegraphics[width=0.95\textwidth]{{"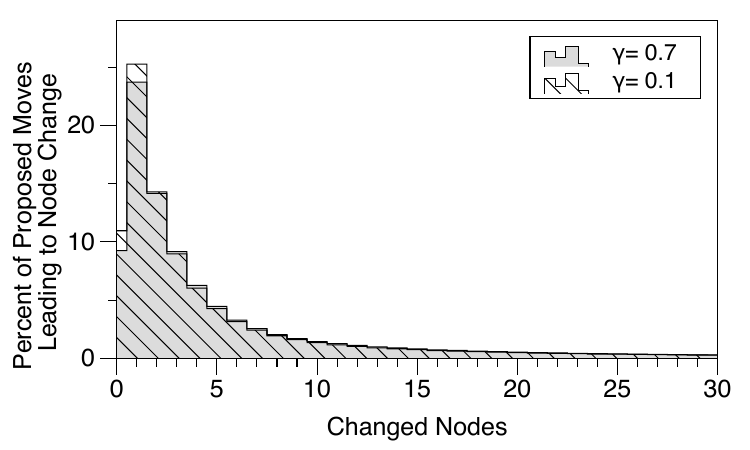"}}
\caption{$\gamma=0.1$ and $\gamma=0.7$}
\label{fig:ChangeNodesHistogram}
\end{subfigure}
 \caption{
 The first three figures show the acceptance probability (median and quartiles) as a function of the relative population change of the proposed move for three values of $\gamma$ with the isoperimetric score turned on to target the measures presented in Fig~\ref{fig:isoperimetric}. On these plots,  we also show histograms of the frequency of proposals with different relative population changes. In the last figure, we plot the histogram of the number of nodes proposed to be swapped when the acceptance probability is positive.
 }
\label{fig:DeltaChange}
\end{figure}

The results of our numerical experiments are shown in Figure~\ref{fig:DeltaChange}. The first three graphs give the distribution of the relative population change in the districts for proposals that change the partition for $\gamma$ equal to $0.1$, $0.4$, and $0.7$. The blue lines in these graphs denote the median acceptance rate for each relative change in populations. The light blue shaded region marks the range of the middle quartiles for the same acceptance rates. The light gray histograms give the percentage of proposals with positive acceptance ratios for each range of relative population change. When $\gamma=0.1$, the acceptance probabilities are reasonable, typically greater than 0.4, even for large relative population changes, this aligns with the relative success of Metropolized Forest Recombination for small $\gamma$ since it typically makes proposals that have a large relative population change. As $\gamma$ increases, the acceptance probability for large relative population changes decreases to zero. By  $\gamma=0.7$, the acceptance probability for even moderate relative population changes decreases to close to zero. This gives insight into why Metropolized Forest Recombination has difficulty mixing at  $\gamma=0.7$ as most of its proposals will have large relative population changes. Yet Metropolized Cycle Walk is still able to make progress as it regularly proposes small population changes that are regularly accepted even at  $\gamma=0.7$ and from time to time larger moves. In this way, Metropolized Cycle Walk can mix better than Metropolized Forest Recombination.
Even at  $\gamma=0.7$, Cycle Walk is regularly proposing much bigger moves relative to a Single Node Flip proposal. Furthermore, because the proposals are balanced, there is less chance that the next proposal for a given pair of districts will just undo the previous move to restore population balance, a phenomenon which is reported when single-node flip proposals are used.
Notice that at $\gamma=0.7$, the median acceptance probability begins to be small even for the more typically proposed relative population changes. This is a possible explanation of why the  Metropolized Cycle Walk seems to mix poorly at even higher $\gamma$.

The fourth figure gives the frequency of the number of nodes changed in proposals with positive acceptance probabilities when $\gamma=0.1$ and $\gamma=0.7$. The two histograms are almost identical which implies that the target measure does not substantively influence the distribution of the number of nodes in the proposal.   We also examined values of $\gamma$ in between the values shown. The plots interpolate between those that are shown.

\subsection{Weighted Graphs}\label{sec:weights}

Up to this point, our numerical investigations have taken place on target measures that do not weight the edges of the graph. However, as explained above in Sections~\ref{sec:cycleWalk} and~\ref{sec: Implemented}, our results generalize to weighted graphs and we conclude our numerical exploration by examining how weighted graphs may contribute to county preservation. The idea of using weighted graphs to preserve counties was originally used with RECOM on minimal spanning trees \cite{colorado}; in this setting, the invariant measure on the space is unknown, but if one switches to a random spanning forests with probabilities proportional to products of edge weights, then the invariant measure has a closed form (again as we present in the above sections). 

In this exploration, we take the dual graph on the South Carolina precincts. We let the edges of the graph be 1 if the two nodes do not share a county and up-weight them, by choosing a number greater than 1, if they are in the same county. This upweight makes it less likely to select trees that have multiple edges that span across counties and will result in districts that keep counties (or pieces of counties) together.  As already mentioned in Section~\ref{sec: Implemented}, we take the  $\alpha$ weights in the \treeOne Cycle Walk equal to the $\beta$  weights in the \treeTwo Cycle Walk in our experiments.

\begin{figure}
\centering
 \includegraphics[width=0.6\textwidth]{{"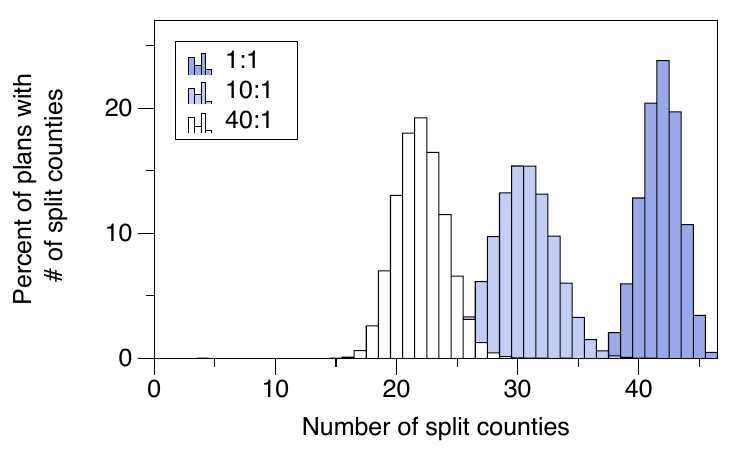"}}
\caption{We show the histograms of the number of split counties in an ensemble of 50-district plans in South Carolina. We up-weight edges that have two nodes in the same county by a factor of 1, 10 and 40 to demonstrate the effect of county splitting on sampling from the weighted spanning forests.}
\label{fig:sccountysplitting}
\end{figure}

We sample the 50 districts of the South Carolina State Senate using the 2020 census population with within-county weights at 1, 10, and 40. We take a 2-step Cycle Walk with 10\% probability and a 1-step Cycle Walk with 90\% probability and run 4 independent chains for each weighting scheme for $10^9$ total steps. Each run took roughly 16 hours of wall clock time on an AMD Ryzen 9 5950X 16-Core Processor. We confirm convergence with the same methods using the 2020 presidential voting data for the general election. We then plot the number of counties that have been split in each case and show the result in \Cref{fig:sccountysplitting}. There are 46 counties in South Carolina. When we run with uniform weights, we find that some of the 50-district plans split all 46 counties. However, as we increase the weight, fewer and fewer counties are split. We remark that several counties in South Carolina are significantly bigger than a single State Senate district and thus must be split.

This study demonstrates the efficacy of the Cycle Walk algorithm at sampling weighted spanning trees that  align with certain policy considerations in redistricting. Of course, encoding policy considerations in terms of weights might not be the most transparent from a policy perspective. It might make more sense to simply provide an energy that reduces the probability in the target measure of configurations with too many split counties \cite{herschlag2020quantifying,RuchoVCC}. The edge weighting presented here could be plausibly used in a proposal chain that then Metropolizes a measure that uses a more transparent measure from a policy perspective.  If one wants strict county preservation, it is often best to use a hierarchical sampling scheme as in \cite{autry2021metropolized,autry2020multiscale}. 

\section{Up-Down Walks and the Balanced Tree Walk}\label{sec:updown}

In addition to mixtures of the $k$-Cycle Walks studied above, there are many other intriguing algorithms to study in the extended tree space. One such idea is a variation on the basic Up-Down Walk presented in \cite{AnariVinzantVuong2021}.
In the context of spanning trees or spanning forests, the basic Up-Down Walk proceeds by adding an edge to a spanning forest with $d$ trees. Typically, the edge is chosen uniformly at random from among the edges not in the spanning forest. The new edge either creates a cycle, thereby making one of the trees no longer a tree, or reduces the number of trees by connecting two trees to produce a single bigger tree. This is the ``up'' step as the number of trees is decreased to $d-1$ by adding an edge. Next, one chooses an edge uniformly at random from among the edges that, if removed, would return the graph to a spanning forest of $d$ trees. This is the ``down'' step as it increases the number of trees from $d-1$ to $d$ by removing an edge. It is straightforward to see that the Up-Down Walk $Q_\downUp$ is symmetric in that for any $\tau$ and $\tau'$ in $\mathcal{F}_d$ one has $Q_\downUp(\tau,\tau')=Q_\downUp(\tau',\tau)$.  Using a standard detailed balance argument, this implies that the uniform measure on $\mathcal{F}_d$ is an invariant measure for $Q_\downUp$. The Up-Down walk is also an irreducible Markov chain; hence, the uniform measure is the only invariant measure. The Up-Down Walk is a related walk that increases the number of trees to $d+1$ by first removing an edge and then adding an edge to return to $d$ trees. 

Both of these walks on spanning forests have been studied extensively. In a recent sequence of seminal works \cite{AnariVinzantVuong2021,Anari_Dereziński_Vuong_Yang_2022},  it was shown that both chains are rapidly mixing.  In this context, rapidly mixing means that for any $\epsilon>0$ the number of steps needed from an arbitrary initial state for the Markov chain to be within $\epsilon$ of its stationary measure in the total-variation distance scales like a polynomial in the system size as measured by the number of vertices.

In a more recent important work \cite{cannon2024samplingbalancedforestsgrids}, the authors show that the probability that a uniform draw from spanning trees on classes of planar graphs can be split into $d$ balanced components is bounded away from zero as the size of the graph grows. An ever-refining, square lattice is one such family of graphs.  This proves a conjecture made in \cite{charikar2023complexitysamplingredistrictingplans} and provides an interesting algorithm for uniformly drawing an element of $\mathcal{F}_d$ that has balanced population.

It surprised many people that a uniformly chosen spanning tree could often be split into $d$ balanced components. Similarly, it might be surprising that when combining two trees with a pair of edges, one often can find another pair of edges to remove that again produces a pair of trees that is again balanced but quite different from the original pair. This fact is central to the effectiveness of the Cycle Walk we have explored in this note.  One nice feature of the Up-Down walk is that its stationary measure is known. This is not the case for the Cycle Walk. Though this is not a large issue, as the Cycle Walk can be Metropolized in a computationally efficient way. As our primary use case employs flexible measures and requires Metropolization, this is not a large downside. However, for theoretical investigations, it complicates issues.

Hence, it would be interesting to find a walk that might share many of the properties of the Cycle Walk but still have a known invariant measure and preserve some of the structure that made the Up-Down walk amenable to analysis.  A small modification of the Up-Down procedure, which we call the Balanced Tree Up-Down Walk, is one of likely many such intermediaries. It acts on the space of spanning trees on the graph that can be split into $d$ balanced trees by removing $d-1$ edges. One adds an edge to the tree uniformly at random among the edges that are not in the tree. One then removes an edge uniformly at random among those that would again result in a single tree that could be split into $d$ balanced subtrees. Throughout this discussion, we could use different ideas of ``balanced.'' For example, we could mean ``strictly'' balanced, where each subtree has the same population or number of vertices.  When only aproximate balance is enforced, finding which edges could be removed from a long loop to produce a new tree which can be could be split in to $d$ approximately balanced districts by removing $d-1$ edges would required traversing a search space that grows exponentially  with $d$ and the loop length. We could choose to pay this computational cost or we could come up with a heuristic for balance that would incur less computational cost.

As defined, the Balanced Tree Up-Down Walk will remain in the space of trees that can be cut into $d$ balanced pieces.  It is straightforward to see that this chain again is symmetric, so that it has a stationary measure which is uniform on all of the trees that can be cut into $d$ balanced parts.  Unlike the simple Up-Down Walk, it is not clear if this Markov chain is irreducible or how to apply the methods used to analyze the original Up-Down chain. Nonetheless, it seems a useful chain to consider, especially for more theoretical questions. We plan to explore this in later work. For more theoretical calculations, it might be useful to add $k$ edges and then remove $k$ edges to return to a balanced tree. This would more likely grant irreducibility and add connectivity in the phase space.

Algorithmically, the above algorithm would impose a ratio between $k$-Cycle walks, as adding an edge would create a loop between 1,2,3... districts that would then be cut to create a new `cuttable' tree. However, instead of adding $k$ edges as we do in the $k$ Cycle Walk (to create a loop across $k$ districts), we would just add 1 edge at each step, and the other $k-1$ edges in the loop would already be present as part of the state space. If we are in the case of exact balance, then a cuttable tree uniquely determines a partition; otherwise, we can further extend the state space to `mark' $d-1$ edges of the tree that we would remove to obtain a unique partition. In the former case, one thing to note about this walk is that because it samples uniformly on cuttable-trees, it samples from a novel measure on the partitions space: The number of cuttable trees associated with a partition is given as the product of the number of trees within each partition multiplied by the number of trees on the \emph{district} graph, where the district graph is given by the multigraph in which we collapse each district to a single node and keep the edges as multiedges between the districts.

More practically, we could perform combinations of 1-Tree Cycle Walks and \emph{tree}-Cycle Walks in which the latter algorithm, inspired by the above structure, first generates a tree on the district graph, then a random edge is added to the tree on the district graph. The new edge will create a cycle of $k$ districts, and we can then perform a $k$-Cycle Walk step.

\subsection{A Symmetric Multi-Edge Extension}

Another similar construction to the Cycle and Up-Down walks may be useful in exploring the theoretical properties of this family of Markov chains. The idea is to use a similar proposal method of adding random edges while still preserving the balance and partition structure by restricting to local choices for choosing which edges to remove. Beginning with a spanning forest with $d$ trees, we could instead choose a pair of edges uniformly at random from all edges not in the spanning forest. There are five possibilities for the structure of the tree with the addition of these edges, and for each, we can specify a set of choices to return to $\mathcal{F}_d$. The two edges could create two edge-independent cycles, corresponding to two independent 1-Tree Cycle Walk steps, in which case one edge on each cycle can be removed uniformly at random. The two edges could both connect the same pair of trees, in which case two edges on the created cycle can be removed, corresponding to a 2-Tree Cycle Walk step. One cycle could be formed by adding an edge within a tree in the forest and the other edge between two trees, in which case an edge on the cycle can be removed uniformly, along with one edge from the joined trees. If both edges connect different pairs of trees, no cycles are formed, and one edge each can be removed from the pairs of trees that were joined; in this case, it is possible that the two added edges will create a tree comprised of three (previously separated) trees, in which case the edge selections are mutually dependent if we seek to sample balanced trees. Finally, in the case that the added edges create dependent cycles that share at least one edge, the proposal is rejected, and the Markov chain remains at the current forest. 

By restricting the choices of edges to remove to components that are impacted by the additions, this preserves more locality and balance than the regular Up-Down walk and also provides a stronger connection to the Cycle Walk, which is a special case of this Markov chain where we only take the first two types of steps. This Markov chain has symmetric transition probabilities since the initial choices of edges to add are uniform and the cases are disjoint, so this modification of the Cycle Walk has the uniform distribution as its stationary measure. 
The balanced condition could also be imposed on each of these choices of edges to remove. This simplifies some of the analysis since edges that are added between distinct trees have to be removed in order to maintain the balance.  Generalizing this procedure to adding $k$ uniform edges at each step is also a natural object for future study. As mentioned above, this would extend the types of moves that are allowed: when at least three edges are added, cycles could be created that connect more than two trees in the original forest, leading to a different type of global move than those available to the Cycle Walk. 

\section{Conclusions and Future Work}

The Cycle Walk presented here offers several advantages over previous MCMC methods for sampling connected graph partitions with application to analyzing redistricting plans. Combining moves at both small and large scales while also efficiently preserving approximate balance and straightforward Metropolization allows the Cycle Walk to sample effectively from a broad class of relevant distributions. Our experiments in Section \ref{sec:NumExp} show that the Cycle Walk can sample efficiently on real-world graphs for values of $\gamma$ that are difficult for previous state-of-the-art methods, including the important case where a non-tree-based compactness metric is incorporated into the energy function. Even in the $\gamma=0$ case that is most favorable for the Forest Recombination method the Cycle Walk is very competitive in terms of the effective sample size while also not requiring sampling a uniform spanning tree at each step, which is expensive from a computational perspective.  Figures \ref{fig:convergenceWGammaStudy}, \ref{fig:ess}, and \ref{fig:CTconvergence} highlight these advantages, while the results of Section \ref{sec:weights} show that the Cycle Walk can also be extended to incorporate weighted graphs, further broadening the sets of policies that can be evaluated.   The efficient implementation described in Section \ref{sec:codeBase} will also  provide practitioners with the ability to better understand the landscape of potential plans and policies by reducing the time it takes to collect a representative ensemble. 

The success of the Cycle Walk in our experiments also suggests some additional avenues for exploration, including investigating the relationship between the properties of the trees observed by the sampler and the corresponding target distributions, as well as combinatorial descriptions of the Markov chain itself. Section \ref{sec:structureCycleWalk} provides some initial experiments along these lines, measuring properties of the proposals at the level of partitions, but similar measurements could also be made on the forests themselves, for example by tracking the lengths of the cycles created by the proposals or the graph-theoretic properties of the trees in each part, which could offer insight into the relationship between the combinatorial structure and acceptance probabilities. Correlations from these types of experiments could then be used to separately Metropolize the 1-Tree Cycle steps to preferentially sample from trees more likely to be cuttable or with other desirable properties. From a theoretical perspective, in addition to exploring the relationships between this walk and those presented in Section \ref{sec:updown} it would be interesting to analyze the chain where the 1-Tree Cycle steps are replaced by uniform resampling of the spanning tree, which is the limiting case of the Cycle Walk as the ratio of 1-Tree steps to 2-Tree steps goes to $\infty:1$. 

\subsection*{Acknowledgments}
 This project started in the spring of 2023 but really hit its stride in the fall of 2023 when all three authors were members of the Simons Laufer Mathematical Sciences Institute (SLMath) program on “Algorithms, Fairness, and Equity.” Thus, this material is based upon work supported by the National Science Foundation under Grant No. DMS-1928930 while the authors were in residence at the Mathematical Sciences Research Institute in Berkeley, California, during the Fall 2023 semester. The stay at SLMath was also supported by the Simons Foundation (G-2021-16778). JCM and GJH would like to acknowledge the support of a grant from Imagine NC First, which supported this and other work around redistricting. All three authors would like to acknowledge the American Institute of Mathematics (AIM), and the NSF for its support of AIM, for a week-long stay during the spring of 2025, when another push on this project occurred. Lastly, the authors acknowledge the use of Claude Code Sonnet 4.6 and Codex GPT-5.5 to identify possible mathematical typos and inconsistencies in the draft manuscript.

\appendix

\section{Tuning Compactness Weights}
\label{apdx:tuning_compactness_weights}
\begin{figure}
\centering
\includegraphics[width=0.5\textwidth]{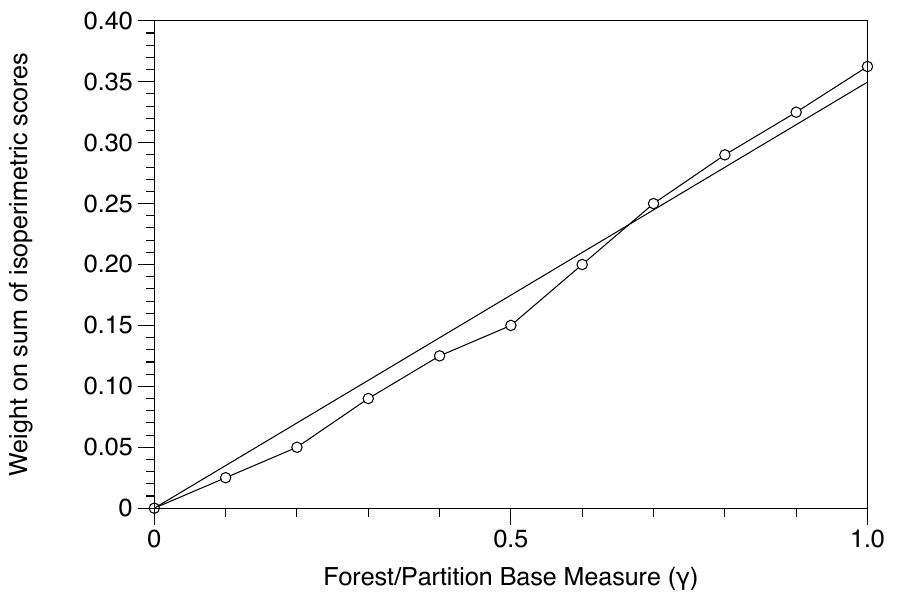}
\caption{We show the multiplicative weight on the sum of isoperimetric scores as a function of $\gamma$ to arrive at the results shown in \Cref{fig:isoperimetric}. We also show the best fit line passing through the origin.}
\label{fig:cweightvgammanc}
\end{figure}

We report the weights used to determine the results displayed in \Cref{fig:isoperimetric}. We find (roughly) linear scaling between $\gamma$ and $c_\gamma$. Fitting a line through the origin, we find a slope of about 0.35. We display the scaling and weights in \Cref{fig:cweightvgammanc}.

\section{The code base}\label{sec:codeBase}
We have implemented the Cycle Walk algorithm that mixes 1-Tree Cycle Walks with 2-Tree Cycle walks in the  Julia\footnote{See \url{https://julialang.org/}} programming language. As already mentioned, the implementation uses splay tree structures with the linked-cut tree structures to store the spanning forests and efficiently implement the needed operations. The code and examples from this paper can be found at \url{https://github.com/jonmjonm/CycleWalk.jl.git}. The Cycle Walk code has also been registered as an official Julia package which can be installed using the Julia package manager. It should be simple for readers to run the code on a graph of their own choosing. Examples of how to run the code can be found at the Quantifying Gerrymandering Project Documentation site \url{https://quantifyinggerrymandering.pages.oit.duke.edu/codedoc/}.  The graphs representing the precincts or census blocks are described using the same JSON format that GerryChain\footnote{See \url{https://mggg.github.io/GerryChain/index.html}} uses which was modified from the Python NetworkX\footnote{See \url{https://networkx.org/}} package. Most of the figures in the paper were made with DataGraph \url{https://www.visualdatatools.com/DataGraph/}.

\bibliographystyle{plain}
\bibliography{ref.bib}

@article{russo2018linking,
  title={Linking and cutting spanning trees},
  author={Russo, Lu{\'\i}s MS and Teixeira, Andreia Sofia and Francisco, Alexandre P},
  journal={Algorithms},
  volume={11},
  number={4},
  pages={53},
  year={2018},
  publisher={MDPI}
}

@misc{cannon2022spanning,
  title =	 {Spanning tree methods for sampling graph partitions},
  author =	 {Sarah Cannon and Moon Duchin and Dana Randall and
                  Parker Rule},
  year =	 2022,
  eprint =	 {2210.01401},
  archivePrefix ={arXiv},
  primaryClass = {physics.soc-ph}
}

@article{fifield2020automated,
  title =	 {Automated redistricting simulation using Markov
                  chain Monte Carlo},
  author =	 {Fifield, Benjamin and Higgins, , Michael and Imai,
                  Kosuke and Tarr, Alexander},
  journal =	 {Journal of Computational and Graphical Statistics},
  volume =	 29,
  number =	 4,
  pages =	 {715--728},
  year =	 2020,
  publisher =	 {Taylor \& Francis}
}

@article{autry2021metropolized,
  title =	 {Metropolized multiscale forest recombination for
                  redistricting},
  author =	 {Autry, Eric A and Carter, Daniel and Herschlag,
                  Gregory J and Hunter, Zach and Mattingly, Jonathan
                  C},
  journal =	 {Multiscale Modeling \& Simulation},
  volume =	 19,
  number =	 4,
  pages =	 {1885--1914},
  year =	 2021,
  publisher =	 {SIAM}
}

@article{mccartan2020sequential,
  title =	 {Sequential Monte Carlo for sampling balanced and
                  compact redistricting plans},
  author =	 {McCartan, Cory and Imai, Kosuke},
  journal =	 {arXiv preprint arXiv:2008.06131},
  year =	 2020
}

@article{deford2019recombination,
  title =	 {Recombination: A family of Markov chains for
                  redistricting},
  author =	 {DeFord, Daryl and Duchin, Moon and Solomon, Justin},
  journal =	 {Harvard Data Science Review},
  volume =	 3,
  number =	 1,
  year =	 2021
}

@article{autry2020multiscale,
  title =	 {Multi-Scale Merge-Split Markov Chain Monte Carlo for
                  Redistricting},
  author =	 {Eric A. Autry and Daniel Carter and Gregory
                  Herschlag and Zach Hunter and Jonathan C. Mattingly},
  year =	 2020,
  eprint =	 {2008.08054},
  archivePrefix ={arXiv},
  primaryClass = {math.PR}
}

@inproceedings{wilsonGeneratingRandomSpanning1996,
  address =	 {{Philadelphia, Pennsylvania, United States}},
  title =	 {Generating Random Spanning Trees More Quickly than
                  the Cover Time},
  isbn =	 {978-0-89791-785-8},
  language =	 {en},
  booktitle =	 {Proceedings of the Twenty-Eighth Annual {{ACM}}
                  Symposium on {{Theory}} of Computing - {{STOC}} '96},
  publisher =	 {{ACM Press}},
  doi =		 {10.1145/237814.237880},
  author =	 {Wilson, David Bruce},
  year =	 1996,
  pages =	 {296-303}
}

@article{Bangia17,
  Author =	 {Sachet Bangia and Christy Vaughn Graves and Gregory
                  Herschlag and Han Sung Kang and Justin Luo and
                  Jonathan C. Mattingly and Robert Ravier},
  Journal =	 {arXiv},
  Title =	 {Redistricting: Drawing the Line},
  Volume =	 {1704.03360},
  Year =	 2017
}

@article{fifield2015,
  Author =	 {Fifield, Benjamin and Higgins, Michael and Imai,
                  Kosuke and Tarr, Alexander},
  Journal =	 {Work. Pap., Princeton Univ., Princeton, NJ},
  Title =	 {A new automated redistricting simulator using
                  {M}arkov chain {M}onte {C}arlo},
  Year =	 2015
}

@article{deford2019redistricting,
  title =	 {Redistricting reform in Virginia: Districting
                  criteria in context},
  author =	 {DeFord, Daryl and Duchin, Moon},
  journal =	 {Virginia Policy Review},
  volume =	 12,
  number =	 2,
  pages =	 {120--146},
  year =	 2019
}

@article{najt2021empirical,
  title =	 {Empirical sampling of connected graph partitions for
                  redistricting},
  author =	 {Najt, Elle and DeFord, Daryl and Solomon, Justin},
  journal =	 {Physical Review E},
  volume =	 104,
  number =	 6,
  pages =	 064130,
  year =	 2021,
  publisher =	 {APS}
}

@misc{herschlag2020nonreversiblemarkovchainmonte,
  title =	 {Non-reversible Markov chain Monte Carlo for sampling
                  of districting maps},
  author =	 {Gregory Herschlag and Jonathan C. Mattingly and
                  Matthias Sachs and Evan Wyse},
  year =	 2020,
  eprint =	 {2008.07843},
  archivePrefix ={arXiv},
  primaryClass = {stat.CO},
  url =		 {https://arxiv.org/abs/2008.07843},
}

@misc{chuang2024multiscaleparalleltemperingfast,
  title =	 {Multiscale Parallel Tempering for Fast Sampling on
                  Redistricting Plans},
  author =	 {Gabriel Chuang and Gregory Herschlag and Jonathan
                  C. Mattingly},
  year =	 2024,
  eprint =	 {2401.17455},
  archivePrefix ={arXiv},
  primaryClass = {physics.soc-ph},
  url =		 {https://arxiv.org/abs/2401.17455},
}

@misc{jonasBlogPost,
  Author =	 {Jonas Eichenlaub},
  Howpublished =
                  {https://sites.duke.edu/quantifyinggerrymandering/2023/08/23/comparing-algorithms-for-generating-ensembles-to-detecting-gerrymandering/},
  Title =	 {{C}omparing {A}lgorithms for {G}enerating
                  {E}nsembles to {D}etecting {G}errymandering},
  Year =	 2023
}

@article{autry2023metropolized,
  title =	 {Metropolized forest recombination for monte carlo
                  sampling of graph partitions},
  author =	 {Autry, Eric and Carter, Daniel and Herschlag,
                  Gregory J and Hunter, Zach and Mattingly, Jonathan
                  C},
  journal =	 {SIAM Journal on Applied Mathematics},
  volume =	 83,
  number =	 4,
  pages =	 {1366--1391},
  year =	 2023,
  publisher =	 {SIAM}
}

@article{MattinglyVaughn2014,
  Adsurl =	 {https://arxiv.org/abs/1410.8796},
  Archiveprefix ={arXiv},
  Author =	 {{Mattingly}, J.~C. and {Vaughn}, C.},
  Eprint =	 {1410.8796},
  Journal =	 {ArXiv e-prints},
  Month =	 oct,
  Primaryclass = {physics.soc-ph},
  Title =	 {{Redistricting and the Will of the People}},
  Year =	 2014
}

@incollection{zhao2022mathematically,
  title =	 {Mathematically quantifying non-responsiveness of the
                  2021 Georgia Congressional districting plan},
  author =	 {Zhao, Zhanzhan and Hettle, Cyrus and Gupta, Swati
                  and Mattingly, Jonathan Christopher and Randall,
                  Dana and Herschlag, Gregory Joseph},
  booktitle =	 {Equity and Access in Algorithms, Mechanisms, and
                  Optimization},
  pages =	 {1--11},
  year =	 2022
}

@article{herschlag2020quantifying,
  title =	 {Quantifying gerrymandering in north carolina},
  author =	 {Herschlag, Gregory and Kang, Han Sung and Luo,
                  Justin and Graves, Christy Vaughn and Bangia, Sachet
                  and Ravier, Robert and Mattingly, Jonathan C},
  journal =	 {Statistics and Public Policy},
  volume =	 7,
  number =	 1,
  pages =	 {30--38},
  year =	 2020,
  publisher =	 {Taylor \& Francis}
}

@article{jcmReport,
  title =	 {Expert Report for {C}ommon {C}ause v. {L}ewis},
  author =	 {Jonathan C. Mattingly},
  journal =	 {{C}ommon {C}ause v. {L}ewis},
  year =	 2019
}

@article{jcmReportHarperVHallMoore,
  title =	 {Expert Report for {H}arper v. {H}all/{M}oore},
  author =	 {Jonathan C. Mattingly},
  journal =	 {{H}arper v. {H}all/{M}oore},
  year =	 2021
}

@book{RuchoVCC,
  title =	 {\textup{Rucho v. Common Cause, No. 18-422, 588
                  U.S. \_\_\_ (2019)}},
  year =	 {}
}

@article{duchinPAreport,
  title =	 {Outlier analysis for Pennsylvania congressional
                  redistricting},
  author =	 {Moon Duchin},
  journal =
                  {https://www.governor.pa.gov/wp-content/uploads/2018/02/md-report.pdf}
}

@misc{najt2019complexitygeometrysamplingconnected,
  title =	 {Complexity and Geometry of Sampling Connected Graph
                  Partitions},
  author =	 {Elle Najt and Daryl DeFord and Justin Solomon},
  year =	 2019,
  eprint =	 {1908.08881},
  archivePrefix ={arXiv},
  primaryClass = {cs.CC},
  url =		 {https://arxiv.org/abs/1908.08881},
}

@misc{cannon2024samplingbalancedforestsgrids,
  title =	 {Sampling Balanced Forests of Grids in Polynomial
                  Time},
  author =	 {Sarah Cannon and Wesley Pegden and Jamie
                  Tucker-Foltz},
  year =	 2024,
  eprint =	 {2310.15152},
  archivePrefix ={arXiv},
  primaryClass = {cs.DM},
  url =		 {https://arxiv.org/abs/2310.15152},
}

@misc{charikar2023complexitysamplingredistrictingplans,
  title =	 {On the Complexity of Sampling Redistricting Plans},
  author =	 {Moses Charikar and Paul Liu and Tianyu Liu and
                  Thuy-Duong Vuong},
  year =	 2023,
  eprint =	 {2206.04883},
  archivePrefix ={arXiv},
  primaryClass = {cs.DS},
  url =		 {https://arxiv.org/abs/2206.04883},
}

@misc{anari2021logconcavepolynomialsivapproximate,
  title =	 {Log-Concave Polynomials IV: Approximate Exchange,
                  Tight Mixing Times, and Near-Optimal Sampling of
                  Forests},
  author =	 {Nima Anari and Kuikui Liu and Shayan Oveis Gharan
                  and Cynthia Vinzant and Thuy Duong Vuong},
  year =	 2021,
  eprint =	 {2004.07220},
  archivePrefix ={arXiv},
  primaryClass = {cs.DS},
  url =		 {https://arxiv.org/abs/2004.07220},
}

@inproceedings{AnariVinzantVuong2021,
  author =	 {Anari, Nima and Liu, Kuikui and Gharan, Shayan Oveis
                  and Vinzant, Cynthia and Vuong, Thuy-Duong},
  title =	 {Log-concave polynomials IV: approximate exchange,
                  tight mixing times, and near-optimal sampling of
                  forests},
  year =	 2021,
  isbn =	 9781450380539,
  publisher =	 {Association for Computing Machinery},
  address =	 {New York, NY, USA},
  url =		 {https://doi.org/10.1145/3406325.3451091},
  doi =		 {10.1145/3406325.3451091},
  booktitle =	 {Proceedings of the 53rd Annual ACM SIGACT Symposium
                  on Theory of Computing},
  pages =	 {408–420},
  numpages =	 13,
  location =	 {Virtual, Italy},
  series =	 {STOC 2021}
}

@article{Anari_Dereziński_Vuong_Yang_2022,
  title =	 {Domain Sparsification of Discrete Distributions
                  Using Entropic Independence},
  volume =	 215,
  rights =	 {Creative Commons Attribution 4.0 International
                  license, info:eu-repo/semantics/openAccess},
  ISBN =	 9783959772174,
  ISSN =	 {1868-8969},
  DOI =		 {10.4230/LIPICS.ITCS.2022.5},
   journal =	 {LIPIcs, Volume 215, ITCS 2022},
  publisher =	 {Schloss Dagstuhl – Leibniz-Zentrum für Informatik},
  author =	 {Anari, Nima and Dereziński, Michał and Vuong,
                  Thuy-Duong and Yang, Elizabeth},
  editor =	 {Braverman, Mark},
  year =	 2022,
  pages =	 {5:1-5:23},
  language =	 {en}
}

@Article{spectralAnalysis, 
  author = 	 {Greg Herschlag and Jeremy L. Marzuola and Jonathan C. Mattingly and Andrew Sun},
  title = 	 {A spectral analysis of tree based Markov chains for redistricting problems},
  journal = 	 {Preprint},
  year = 	 2025
}

@article {Kruskal56,
    AUTHOR = {Kruskal, Jr., Joseph B.},
     TITLE = {On the shortest spanning subtree of a graph and the traveling
              salesman problem},
   JOURNAL = {Proc. Amer. Math. Soc.},
  FJOURNAL = {Proceedings of the American Mathematical Society},
    VOLUME = {7},
      YEAR = {1956},
     PAGES = {48--50},
      ISSN = {0002-9939,1088-6826},
   MRCLASS = {56.0X},
  MRNUMBER = {78686},
MRREVIEWER = {H.\ W.\ Kuhn},
       DOI = {10.2307/2033241},
       URL = {https://doi.org/10.2307/2033241},
}

@misc{babson2024modelsrandomspanningtrees,
      title={Models of random spanning trees}, 
      author={Eric Babson and Moon Duchin and Annina Iseli and Pietro Poggi-Corradini and Dylan Thurston and Jamie Tucker-Foltz},
      year={2024},
      eprint={2407.20226},
      archivePrefix={arXiv},
      primaryClass={cs.DM},
      url={https://arxiv.org/abs/2407.20226}, 
}

@book{Cormen,
    AUTHOR = {Cormen, Thomas H. and Leiserson, Charles E. and Rivest, Ronald
              L. and Stein, Clifford},
     TITLE = {Introduction to algorithms},
   EDITION = {Third},
 PUBLISHER = {MIT Press, Cambridge, MA},
      YEAR = {2009},
     PAGES = {xx+1292},
      ISBN = {978-0-262-03384-8},
   MRCLASS = {68-01 (05-01 05C85 68P05 68P10 68Q25 68Wxx)},
  MRNUMBER = {2572804},
}

@misc{Clelland,
      title={Compactness statistics for spanning tree recombination}, 
      author={Jeanne N. Clelland and Nicholas Bossenbroek and Thomas Heckmaster and Adam Nelson and Peter Rock and Jade VanAusdall},
      year={2021},
      eprint={2103.02699},
      archivePrefix={arXiv},
      primaryClass={physics.soc-ph},
      url={https://arxiv.org/abs/2103.02699}, 
}

@article{colorado,
  title={{Colorado in Context: Congressional Redistricting and Competing Fairness Criteria in Colorado}},
  author={Jeanne Clelland and  Haley Colgate and Daryl DeFord and Beth Malmskog and Flavia Sancier-Barbosa},
  journal={Journal of Computational Social Science},
  year={2021},
month= may,
volume={5},
pages={180-226}
}

@article{SwendsenWang,
  title = {Nonuniversal critical dynamics in Monte Carlo simulations},
  author = {Swendsen, Robert H. and Wang, Jian-Sheng},
  journal = {Phys. Rev. Lett.},
  volume = {58},
  issue = {2},
  pages = {86--88},
  numpages = {0},
  year = {1987},
  month = {Jan},
  publisher = {American Physical Society},
  doi = {10.1103/PhysRevLett.58.86},
  url = {https://link.aps.org/doi/10.1103/PhysRevLett.58.86}
}

\end{document}